\documentclass[english,aps,a4paper,prd,groupedaddress,nofootinbib,preprint,floatfix,preprintnumbers]{revtex4-2}
\usepackage[utf8]{inputenc}
\usepackage[english]{babel}
\usepackage{natbib}
\usepackage{graphicx}
\usepackage{subcaption}
\usepackage{float}
\usepackage{multirow}
\usepackage{amssymb,amsmath,bbm,mathrsfs,enumerate}
\usepackage{color}
\usepackage{hyperref}
\usepackage{cleveref}

\newcommand{\ba}{\begin{eqnarray}}
\newcommand{\ea}{\end{eqnarray}}

\def\OB{\mathcal{O}}

\begin{document}

\title{Combining Evolutionary Strategies and Novelty Detection to go Beyond the Alignment Limit of the $Z_3$ 3HDM}

\author{Jorge Crispim Romão}
\email{jorge.romao@tecnico.ulisboa.pt}
\affiliation{CFTP, Departamento de F\'{i}sica, Instituto Superior T\'{e}cnico,
Universidade de Lisboa,\\
Avenida Rovisco Pais 1, 1049 Lisboa, Portugal}
\author{Miguel Crispim Romão}
\email{miguel.romao@durham.ac.uk}
\affiliation{Institute for Particle Physics Phenomenology, Durham University, Durham DH1 3LE, UK}
\affiliation{LIP -- Laborat\'orio de Instrumenta\c{c}\~ao e F\'isica Experimental de Part\'iculas, Escola de Ciências, Campus de Gualtar, Universidade do Minho, 4701-057 Braga, Portugal}
\today

\begin{abstract}
	We present a novel Artificial Intelligence approach for Beyond the
	Standard Model parameter space scans by augmenting an Evolutionary Strategy with Novelty Detection. Our approach leverages the power of
	Evolutionary Strategies, previously shown to quickly converge to the
	valid regions of the parameter space, with a \emph{novelty reward} to
	continue exploration once converged. Taking the $Z_3$ 3HDM as our
	Physics case, we show how our methodology allows us to quickly explore
	highly constrained multidimensional parameter spaces, providing up to
	eight orders of magnitude higher sampling efficiency when compared
	with pure random sampling and up to four orders of magnitude when
	compared to random sampling around the alignment limit. In turn, this
	enables us to explore regions of the parameter space that have been
	hitherto overlooked, leading to the possibility of novel phenomenological realisations of the $Z_3$ 3HDM that had not been considered before.
\end{abstract}

\preprint{IPPP/24/04}
\preprint{CFTP/24-002}

\maketitle

\section{Introduction}
\label{s:intro}
The Standard Model (SM) of particle physics has demonstrated
remarkable success in accurately describing the electroweak and strong
interactions. However, experimental challenges such as neutrino mass,
dark matter, and the baryonic asymmetry of the Universe
prompt the exploration of physics Beyond the SM (BSM). In many cases,
BSM theories involve expanding the minimal scalar sector of the SM,
characterised by a single Higgs doublet.
Multi-Higgs doublet models are particularly prominent
among these extensions, primarily because they maintain the tree-level
value of the electroweak $\rho$-parameter, in good agreement with
experimental observations.
The extensively studied two Higgs doublet model (2HDM)~\cite{Branco:2011iw}
has provided valuable insights. More recently, there has been a surge
of interest in the investigation of three Higgs doublet models (3HDMs)
\cite{Keus:2013hya,Ivanov:2012fp,Pilaftsis:2016erj,Akeroyd:2016ssd,Das:2019yad,Alves:2020brq,Logan:2020mdz,Chakraborti:2021bpy,Boto:2021qgu,Bento:2022vsb,Boto:2022uwv,Plantey:2022jdg,Das:2022gbm,Kuncinas:2023ycz,Boto:2023nyi}, where the scalar sector
encompasses three Higgs doublets. These models show great promise as
they possess the essential ingredients to address the
challenges posed by dark matter and the baryonic asymmetry of the
Universe.

With no unambiguous signs of New Physics in general and of extra exotic scalars in particular, BSM phenomenology is faced with an ever increasing and ever more restricting list of (direct and indirect) experimental constraints, in addition to any theoretical and self-consistency constraints. From a model building perspective, this means that the allowed regions of BSM parameter spaces are effectively \emph{shrinking}, making finding such regions ever more difficult or outright impractical, even if those regions host points which are very good fits to the data. To mitigate this, BSM phenomenologists often simplify the problem in two possible ways. The first approach is to simplify the constraints and leave some questions unanswered, to be eventually addressed by an Ultra Violet completion. The second approach relies on simplifying the sampling space by changing the priors from which the parameters are drawn, usually by restricting the parameter space to a subregion where known valid points had been found before or for which there is a limiting case, such as the case of the alignment limit in multi Higgses models. While in the former case one ends up with an incomplete model, in the latter case one is left with the worrisome prospect of missing out possible phenomenological signatures.

In recent years, Artificial Intelligence (AI) in general, and Machine Learning (ML) in particular, have received considerable attention by the High Energy Physics (HEP) community with a wide range of applications~\cite{Feickert:2021ajf}. Of particular interest to this work are the ongoing attempts to explore the parameter space of highly constrained and multidimensional BSM scenarios, where sampling points that respect theoretical and experimental bounds poses a great challenge. The first attempts to mitigate this problem using AI/ML are based on using supervised classifiers and regressors to produce estimates of BSM predictions of physical observables and physical quantities~\cite{Caron:2016hib,Ren:2017ymm,Staub:2019xhl,Kronheim:2020vct,Hammad:2022wpq}, bypassing the computational overhead often associated with these. Other approaches leverage the active learning framework to track the ground truth of the observables to draw the boundary of the allowed regions of the parameter space~\cite{Caron:2019xkx,Goodsell:2022beo}. However, we point out that the usage of AI/ML for BSM parameter spaces studies is not restricted to exploration, as generative models have been studied to provide a possible way of replicating points from valid regions~\cite{Hollingsworth:2021sii} or to gain new insights through latent space visualisation~\cite{Baretz:2023mra}.

The approaches cited above rely heavily on the quality of the trained ML model, namely on the coverage of the parameter space provided by the points used, especially in or near valid regions. A different approach was presented in~\cite{deSouza:2022uhk}, where different AI-based exploration algorithms were used to explore the parameter space of the cMSSM (four parameters) and the pMSSM (16 parameters) by reframing the problem as a black-box optimisation problem. In such an approach, exploration starts with just a few random points from which iterative algorithms will progressively suggest better points through trial and error. Although the Physics cases in~\cite{deSouza:2022uhk} were not especially realistic, as only Higgs mass and Dark Matter relic density constraints were used, the methodology provided orders of magnitude of sampling efficiency improvement over random sampling, while still capturing a global picture of the parameter space.

In this paper, we will extend and build on top of that of~\cite{deSouza:2022uhk}. In that work, an Evolutionary Strategy algorithm was observed to \emph{eagerly} converge to the valid region of the parameter space and to stop exploring once converged. To mitigate this, in~\cite{deSouza:2022uhk} the algorithm was endowed with a restart strategy to draw a more global picture of the valid region of the parameter space. In this work, we will take a different approach by incorporating a \emph{novelty reward} into the black-box optimisation problem to drive the exploration algorithm away from the regions already explored. As we will see, our approach retains the benefits of drastically improving sampling efficiency while gaining a novel approach to charter the valid region of the parameter space.

Although the proposed approach is general to any BSM parameter space,\footnote{In fact, the methodology presented in this paper is applicable to \emph{any} sampling in highly constrained multidimensional spaces, not only to BSM phenomenology.} we will use our novel methodology to perform a realistic search on the highly constrained $Z_3$ 3HDM model, where all known and possible constraints will be considered. This poses a terrific challenge for sampling valid points from the parameter space, which has led previous studies to consider sampling around the alignment limit. Therefore, the $Z_3$ 3HDM provides an ideal scenario not only to develop our methodology, but to use to do new Physics by exploring points beyond random sampling strategies around the alignment limit and to discover novel phenomenological realisations of the $Z_3$ 3HDM that have been obfuscated in the past such strategies.

This paper is organised as follows. In~\cref{s:model} we present the $Z_3$ 3HDM model which is the Physics subject of our study. In~\cref{s:constraints} we present its constraints, both theoretical and experimental. In~\cref{s:scan} we outline the random sampling strategy near the alignment limit, which we will use to compare our methodology. In~\cref{s:AI} we introduce the AI scan strategy, redefining the scan as black-box optimisation, and the \emph{novelty award} based on density estimation. In~\cref{s:results} we present and analyse the results obtained with our methodology, showcasing the versatility of our approach. Finally, in~\cref{s:conclusions} we conclude our discussion and point out novel directions of work.


\section{\label{s:model}The $Z_3$ 3HDM Model}

\subsection{\label{subsec:scalar}Scalar Sector}

For the potential of the $Z_3$ 3HDM model, denoted $V_{Z_3}$, we use
the conventions of~\cite{Das:2019yad}. The $Z_3$-invariant terms, given by $\phi_i\to\phi_i'=(S_{Z_3})_{ij}\phi_j$,
{where
		\begin{equation}
			S_{Z_3}=\text{diag}(1,e^{i\frac{2\pi}{3}}e^{-i\frac{2\pi}{3}}) .
		\end{equation}
	}
can be expressed as:
\begin{equation}\label{Z3potential}
	V_{Z_3}=V_{2}+V_{4},
\end{equation}
with the quartic part represented by:
\begin{align}
	V_{4} & =\lambda_1(\phi_1^\dagger\phi_1)^2 + \lambda_2(\phi_2^\dagger\phi_2)^2 + \lambda_3(\phi_3^\dagger\phi_3)^2 + \lambda_4(\phi_1^\dagger\phi_1)(\phi_2^\dagger\phi_2) + \lambda_5(\phi_1^\dagger\phi_1)(\phi_3^\dagger\phi_3) \nonumber            \\
	      & \quad + \lambda_6(\phi_2^\dagger\phi_2)(\phi_3^\dagger\phi_3) + \lambda_7(\phi_1^\dagger\phi_2)(\phi_2^\dagger\phi_1) + \lambda_8(\phi_1^\dagger\phi_3)(\phi_3^\dagger\phi_1) + \lambda_9(\phi_2^\dagger\phi_3)(\phi_3^\dagger\phi_2) \nonumber \\
	      & \quad + \left[\lambda_{10}(\phi_1^\dagger\phi_2)(\phi_1^\dagger\phi_3) + \lambda_{11}(\phi_1^\dagger\phi_2)(\phi_3^\dagger\phi_2) + \lambda_{12}(\phi_1^\dagger\phi_3)(\phi_2^\dagger\phi_3)+\text{h.c.}\right]. \label{Z3quartic}
\end{align}
The quadratic part, denoted as $V_{2}$, is given by:
\begin{equation}
	V_{2}=m_{11}^2\phi_1^\dagger\phi_1+m_{22}^2\phi_2^\dagger\phi_2+m_{33}^2\phi_3^\dagger\phi_3 +\left[m_{12}^2(\phi_1^\dagger\phi_2) + m_{13}^2(\phi_1^\dagger\phi_3) + m_{23}^2(\phi_2^\dagger\phi_3)+\text{h.c.}\right],
\end{equation}
which includes terms, $m_{12}^2$, $m_{13}^2$, and $m_{23}^2$, responsible for breaking the symmetry softly.

Following spontaneous symmetry breaking (SSB), the three doublets can be parameterised in terms of their component fields as:
\begin{equation}
	\phi_i=\begin{pmatrix} w_k^\dagger \\ (v_i+x_i+i\,z_i)/\sqrt{2}\end{pmatrix} \,\,,\qquad (i=1,2,3)\label{fielddefinitions}
\end{equation}
where $v_i/\sqrt{2}$ corresponds to the vacuum expectation value (vev) for the neutral component of $\phi_i$. It is assumed that the scalar sector of the model explicitly and spontaneously conserves CP. Under this assumption, all parameters in the scalar potential are real, and the vevs $v_1$, $v_2$, $v_3$ are also real.

The scalar potential in~\cref{Z3potential} contains eighteen parameters, and the vevs are parameterised as follows:
\begin{equation}\label{3hdmvevs}
	v_1=v \cos \beta_1 \cos \beta_2\,,\qquad v_2=v \sin \beta_1 \cos \beta_2\, ,\qquad v_3=v \sin \beta_2,
\end{equation}
leading to the Higgs basis
\cite{Georgi:1978ri,Donoghue:1978cj,Botella:1994cs} obtained by the
following rotation,
\begin{equation}\label{higgsbasisZ3}
	\begin{pmatrix} H_0 \\ R_1 \\ R_2 \end{pmatrix}
	=
	\mathcal{O}_\beta
	\begin{pmatrix} x_1 \\ x_2 \\ x_3 \end{pmatrix}
	=
	\begin{pmatrix} \cos{\beta_2} \cos{\beta_1} & \cos{\beta_2} \sin{\beta_1} & \sin{\beta_2} \\ -\sin{\beta_1} & \cos{\beta_1} & 0 \\ -\cos{\beta_1} \sin{\beta_2} & -\sin{\beta_1} \sin{\beta_2} & \cos{\beta_2}\end{pmatrix}
	\begin{pmatrix} x_1 \\ x_2 \\ x_3 \end{pmatrix} .
\end{equation}
%

Orthogonal matrices, denoted as $\textbf{R}$, $\textbf{P}$ and
$\textbf{Q}$, diagonalise the squared mass matrices in the CP-even
scalar, CP-odd scalar, and charged scalar sectors. These matrices are
crucial for transforming the weak basis into the physical mass basis for
states with well-defined masses.
Although this has already been discussed
before~\cite{Das:2019yad,Boto:2021qgu,Boto:2021}, for completeness and
to fix our
notation, we give here the rotations that relate the mass eigenstates to
the weak basis states.

For the neutral scalar sector, the mass terms
can be extracted through the following rotation,
\begin{equation}\label{CPevenDiag}
	\begin{pmatrix} h_1 \\ h_2 \\ h_3
	\end{pmatrix}=\mathcal{O}_\alpha
	\begin{pmatrix} x_1 \\ x_2 \\ x_3
	\end{pmatrix},
\end{equation}
where we take $h_1 \equiv h_{125}$ to the be the 125 GeV Higgs boson
found at LHC.
The form chosen for $\mathcal{O}_\alpha\equiv \textbf{R}$ is
\begin{equation}\label{matrixR}
	\textbf{R}\equiv\mathcal{O}_\alpha=\mathcal{R}_3.\mathcal{R}_2.\mathcal{R}_1 ,
\end{equation}
where the matrices $\mathcal{R}_i$ are
\begin{equation}
	\mathcal{R}_1=
	\begin{pmatrix}
		\cos{\alpha_1} & \sin{\alpha_1} & 0 \\ - \sin{\alpha_1} &
		\cos{\alpha_1} & 0                  \\ 0 & 0 & 1
	\end{pmatrix},\,     \mathcal{R}_2=
	\begin{pmatrix}
		\cos{\alpha_2} & 0 & \sin{\alpha_2} \\ 0 & 1 & 0 \\ -
		\sin{\alpha_2} & 0 &
		\cos{\alpha_2}
	\end{pmatrix},\,     \mathcal{R}_3=
	\begin{pmatrix}
		1              & 0              & 0 \\ 0 &  \cos{\alpha_3} &  \sin{\alpha_3} \\ 0 & -
		\sin{\alpha_3} & \cos{\alpha_3}
	\end{pmatrix}.
\end{equation}

For the charged and pseudoscalar sectors, the physical
scalars can be obtained via the following $3\times 3$ rotations,
\begin{equation}
	\label{eq:CPoddCharged}
	\begin{pmatrix} G^0 \\ A_1 \\A_2
	\end{pmatrix}=\mathcal{O}_{\gamma_1}
	\mathcal{O}_\beta
	\begin{pmatrix} z_1 \\ z_2 \\ z_3
	\end{pmatrix},\qquad
	\begin{pmatrix} G^+ \\ H_1^+ \\ H_2^+
	\end{pmatrix}
	=\mathcal{O}_{\gamma_2} \mathcal{O}_\beta
	\begin{pmatrix}
		w_1^\dagger \\ w_2^\dagger \\ w_3^\dagger
	\end{pmatrix},
\end{equation}
where, the rotation matrices are given by
\begin{equation}
	\label{ogamma1And2}
	\mathcal{O}_{\gamma_1}=
	\begin{pmatrix}
		1 & 0 & 0 \\ 0& \cos{\gamma_1} & -\sin{\gamma_1} \\ 0 & \sin{\gamma_1} & \cos{\gamma_1}
	\end{pmatrix},\qquad
	\mathcal{O}_{\gamma_2}=
	\begin{pmatrix}
		1 & 0 & 0 \\ 0& \cos{\gamma_2} & -\sin{\gamma_2} \\ 0 & \sin{\gamma_2} & \cos{\gamma_2}
	\end{pmatrix} .
\end{equation}

For later use, we define the matrices $\textbf{P}$ and $\textbf{Q}$
as the combinations that connect the weak basis to the physical
mass basis for the CP odd and charged Higgs scalars, respectively,
\begin{equation}\label{matrixP}
	\textbf{P}\equiv\mathcal{O}_{\gamma_1} \mathcal{O}_\beta,
	\qquad
	\textbf{Q}\equiv\mathcal{O}_{\gamma_2} \mathcal{O}_\beta .
\end{equation}

As the states in the physical basis have well-defined masses, we can
obtain relations between the set
\ba
&&\left\{v,\beta_1,\beta_2,m_{h1},m_{h2},m_{h3},m_{A1},m_{A2},m_{H_1^\pm},m_{H_2^\pm},\alpha_1,\alpha_2,\alpha_3,\gamma_1,\gamma_2\right\} ,\label{setphysical}
\ea
and the potential parameters in~\cref{Z3potential},
as shown in Ref.~\cite{Das:2019yad,Boto:2021qgu,Boto:2021}.

\subsection{Higgs-Fermion Yukawa Interactions}

In the Type-I models considered here\footnote{For
	a detailed discussion of all the types of Higgs-Fermion couplings
	that lead to Natural Flavour
	Conservation (NFC) see Ref.\cite{Boto:2022uwv}.}, fermion fields are unaffected by the $Z_3$ transformation, allowing them to couple only to $\phi_3$. The Yukawa couplings to fermions are expressed compactly as:
\begin{equation}\label{eq:couplingNeutralFerm}
	\mathscr{L}_{\rm Y}\ni -\frac{m_f}{v}\bar{f}(a^f_j+i\, b^f_j\gamma_5)fh_j ,
\end{equation}
where $h_j\equiv(h_1,h_2,h_3,A_1,A_2)_j$ represents the physical Higgs
fields.  For
completeness, we list the couplings $a_j^f$ and $b_j^f$ here~\cite{Boto:2022uwv}. We have,
\begin{align}
	a_j^f \to &
	\frac{\textbf{R}_{j,3}}{\hat{v_3}},
	\qquad\qquad j=1,2,3\qquad \text{for all leptons and down quarks}
	,\nonumber                                                                     \\[2pt]
	b_j^f \to &
	\frac{\textbf{P}_{j-2,3}}{\hat{v_3}},
	\qquad\quad j=4,5\quad\qquad \text{for all leptons and down quarks} ,\nonumber \\[2pt]
	a_j^f \to &
	\frac{\textbf{R}_{j,3}}{\hat{v_3}},
	\qquad\qquad j=1,2,3\qquad \text{for all up quarks} ,\nonumber                 \\[2pt]
	b_j^f \to &
	-\frac{\textbf{P}_{j-2,3}}{\hat{v_3}},
	\quad\quad j=4,5\quad\qquad \text{for all up quarks} ,
	\label{eq:coeffNeutralFerm-Type-I}
\end{align}
{where we have defined $\hat{v}_i=v_i/v$}.

For the charged Higgs, $H_1^\pm$ and $H_2^\pm$, the Yukawa couplings to fermions are expressed as:
\begin{eqnarray}
	\mathscr{L}_{\rm Y} &\ni& \frac{\sqrt{2}}{v}\bar{\psi}_{d_i}\left[m_{\psi_{d_i}} V_{ji}^\ast\, \eta_k^L P_L + m_{\psi_{u_j}} V_{ji}^\ast\, \eta_k^R P_R\right] \psi_{u_j} H_k^-
	\nonumber\\
	&&+ \frac{\sqrt{2}}{v}\bar{\psi}_{u_i}\left[m_{\psi_{d_j}} V_{
				ij}\, \eta_k^L P_R + m_{\psi_{u_i}} V_{ij}\, \eta_k^R P_L \right] \psi_{d_j} H_k^+,
	\label{eq:couplingChargedFerm}
\end{eqnarray}
where $(\psi_{u_i},\psi_{d_i})$ is $(u_i,d_i)$ for quarks or
$(\nu_i,\ell_i)$ for leptons. The coefficients $\eta_k^{\ell\,L}$,
$\eta_k^{\ell\,R}$, $\eta_k^{q\,L}$, and $\eta_k^{q\,R}$ are %
\begin{equation}
	\label{eq:coeffChargedFerm-Type-I}
	\eta_k^{\ell\,L}=-\frac{\textbf{Q}_{k+1,3}}{\hat{v_3}}\,,\quad\eta_k^{\ell\,R}=
	0\,,\quad\eta_k^{q\,L} =-\frac{\textbf{Q}_{k+1,3}}{\hat{v_3}}\,,
	\quad\eta_k^{q\,R}=\frac{\textbf{Q}_{k+1,3}}{\hat{v_3}}\,,\quad
	\text{k=1,2}\, .
\end{equation}


\section{Constraints\label{s:constraints}}

In this section, we outline the various constraints necessary to impose theoretical and phenomenological consistency on the model parameters. The specifics of these constraints in the context of 3HDM are well established, as documented in previous works~\cite{Boto:2021qgu, Boto:2022uwv, Boto:2023nyi}. For brevity, we provide a brief list here, deferring further elaboration to~\cref{s:appendix_A}.

From a phenomenological standpoint, our primary objective is to ensure the existence of an SM-like Higgs, identifiable as the scalar boson detected at the LHC. As demonstrated in \cite{Das:2019yad}, achieving this involves staying close to the 'alignment limit,' characterised in the 3HDM by the conditions:
\begin{equation}
	\label{eq:1-cons}
	\alpha_1 = \beta_1 \,, \qquad
	\alpha_2 = \beta_2 \,.
\end{equation}

In this limit, the lightest CP-even scalar, denoted as $h$, exhibits exact SM-like couplings at the tree level, automatically satisfying constraints from Higgs signal strengths. However, our interest lies in exploring permissible deviations from this precise alignment. To this end, we use the signal strength formalism, comparing the results with the experimental limits~\cite{ATLAS:2022vkf}.

Subsequently, we must address constraints stemming from electroweak
precision parameters, specifically $S$, $T$, and $U$. We use the analytic expressions of Ref.\cite{Grimus:2007if}, contrasting them with the fit values provided in Ref.\cite{Baak:2014ora}. Notably, similar to the 2HDM scenario, we can bypass $T$-parameter constraints by imposing~\cite{Das:2022gbm}:
\begin{equation}
	m_{H_1^\pm} = m_{A1} \,, \qquad m_{H_2^\pm} = m_{A2} \,, \qquad \gamma_1 =\gamma_2 \,.
\end{equation}
as we will explain in~\cref{s:scan}.

Additionally, we incorporate constraints arising from flavour data, as
detailed in \cref{s:appendix_A}. Direct searches at the LHC for the heavy non-standard scalars are also considered, employing \texttt{HiggsTools-1.1.3} following Ref.~\cite{Bahl:2022igd}, which provides a comprehensive list of relevant experimental searches.


For theoretical constraints, we insist on the perturbativity of Yukawa
couplings, perturbative unitarity, and BFB (bounded from below)
conditions. The implementation details for these constraints can be found
in~\cref{s:appendix_A}.

\section{Random Scan strategy}
\label{s:scan}

We developed a dedicated code specifically for the $Z_3$ constrained
3HDM~\footnote{The Feynman rules for this
	model were derived with the software
	\texttt{FeynMaster}\cite{Fontes:2019wqh,Fontes:2021iue} and automatically
	imported into the code.},
building upon our earlier
codes~\cite{Fontes:2014xva,Florentino:2021ybj,Boto:2021qgu}. A
thorough exploration of the parameter space was conducted using~\cref{setphysical}. Our fixed inputs remained $v = 246,\text{GeV}$ and
$m_{h1} = 125,\text{GeV}$. Random values were assigned within the
following ranges:
\begin{align}
	 & \alpha_1\,, \alpha_2\,, \alpha_3\,, \gamma_1\,, \gamma_2\, \in
	\left[-\frac{\pi}{2},\frac{\pi}{2}\right];\qquad
	\tan{\beta_1}\,,\tan{\beta_2}\,\in \left[0.3,10\right];
	\nonumber                                                         \\[8pt]
	 & m_{H_1}\equiv m_{h_2}\,, m_{H_2}\equiv m_{h_3}\,
	\in \left[125,1000\right]\,\text{GeV};                            \\[8pt]
	 &
	m_{A_1}\,,m_{A_2},m_{H_1^\pm}\,,m_{H_2^\pm}\,
	\in \left[100,1000\right]\,\text{GeV};                            \\[8pt]
	 &
	m^2_{12},m^2_{13},m^2_{23} \in \left[\pm 10^{-1},\pm 10^{7}\right]\,
	\text{GeV}^2\, ,
	\label{eq:scanparameters}
\end{align}
where the last expression applies only to the soft
masses that are not obtained as derived quantities (see
Ref.\cite{Boto:2021qgu} for the complete expressions).
{Note that $m^2_{ij}$ have dimensions of $(\text{Mass})^2$, but despite
their notation they are not positive definite}.

However, this extensive scan exhibited low efficiency (as detailed in~\cref{tab:efficiency} below). Recognising the significance of
alignment in 3HDM~\cite{Das:2019yad,Boto:2021qgu,Chakraborti:2021bpy},
where alignment is defined by the lightest Higgs scalar having
Standard Model (SM) couplings, we imposed constraints to enhance
efficiency. Initially, aligning $\alpha_1$ with $\beta_1$ and
$\alpha_2$ with $\beta_2$ (\cref{eq:1-cons}) did not yield
enough good points. Ref.~\cite{Chakraborti:2021bpy} introduced an
additional condition :
\begin{equation}
	\label{eq:44}
	\gamma_1=\gamma_2=-\alpha_3,\quad m_{H_1}=m_{A_1}=m_{H_1^\pm},\quad
	m_{H_2}=m_{A_2}=m_{H_2^\pm} .
\end{equation}
This, alongside~\cref{eq:1-cons}, led to a symmetric form of the
quartic part of the
potential\cite{Darvishi:2019dbh,Chakraborti:2021bpy}. In fact,  these conditions
simplified the potential to
\begin{equation}
	\label{eq:1}
	V_{\rm Sym, Lim} = \lambda_{\rm
		SM} \left[ (\phi_1^\dagger \phi_1) + (\phi_2^\dagger \phi_2)
		+(\phi_3^\dagger \phi_3) \right]^2\, ,
\end{equation}
where
\begin{equation}
	\label{eq:2}
	\lambda_{\rm SM} = \frac{m_h^2}{2 v^2} ,
\end{equation}
is the  the SM quartic Higgs coupling. All $\lambda_i$
vanish or are expressed in terms of $\lambda_{\rm SM}$.
The validity of~\cref{eq:1-cons} and ~\cref{eq:44}
also implies that the soft masses can be explicitly
solved, that is, they are no more independent parameters (see
Ref.\cite{Boto:2021qgu} for complete expressions). Now it should
be clear why all such points are good points. Due to
alignment, the LHC results on the $h_{125}$ are easily obeyed, whereas
the perturbativity unitarity, STU and the other constraints are
automatically obeyed.

To facilitate efficiency and consider deviations from strict
alignment, two conditions were introduced~\cite{Boto:2021qgu}. The
first denoted "Al-1", allowed a percentage deviation:
\begin{equation}
	\label{Al-1}
	\frac{\alpha_1}{\beta_1},
	\frac{\alpha_2}{\beta_2} \in\, [0.5,1.5]\, .
	\ \ \ \textbf{(Al-1)}
\end{equation}
The condition "Al-2" was more stringent, combining Al-1 with six additional conditions:
\begin{equation}
	\label{Al-2}
	\frac{\alpha_1}{\beta_1},
	\frac{\alpha_2}{\beta_2},
	\frac{\gamma_2}{\gamma_1},
	\frac{-\alpha_3}{\gamma_1},
	\frac{m_{A_1}}{m_{H_1}},
	\frac{m_{H_1^\pm}}{m_{H_1}},
	\frac{m_{A_2}}{m_{H_2}},
	\frac{m_{H_2^\pm}}{m_{H_2}} \in\, [0.5,1.5]\, .
	\ \ \ \textbf{(Al-2)}
\end{equation}
These conditions, especially Al-2, improved the generation of
meaningful points beyond the SM, even with a departure from strict
alignment.


\section{Artificial Intelligence Black-Box Optimiser Scan Strategy}
\label{s:AI}

To quickly explore the parameter space, we will employ the AI black box optimisation approach to parameter space sampling first presented in~\cite{deSouza:2022uhk}. In this approach, a constraint function, $C$, is introduced,
\begin{equation}
	C(\OB) = \max(0, - \OB + \OB_{LB} , \OB - \OB_{UB})\, ,
\end{equation}
where $\OB$ is the value of an observable (or a constrained quantity), $\OB_{LB}$ is its lower bound (i.e. its lowest allowed value), and $\OB_{UB}$ its upper bound (i.e. its highest allowed value). Here, $\OB$ is obtained by some computational routine that maps the parameter space to physical quantities, where the details of such routine are irrelevant and can therefore be taken as a black-box. If the value of $\OB$ is within its lower and upper bounds, $C(\OB)$ returns $0$, otherwise it returns a positive number that measures \emph{how far} the value of $\OB$ is from its allowed interval. $\OB$ is dependent on the parameters of the model, $\theta=(\alpha_1, \beta_1,\dots)\in \mathcal{P}$ (where $\mathcal{P}$ is the parameter space defined by~\cref{eq:scanparameters}), that is, $\OB=\OB(\theta)$, which implies that $C(\OB)=C(\OB(\theta))=C(\theta)$. Therefore, the set of valid points, $\mathcal{V}$, that satisfy a constraint can be defined in terms of $\theta$ as
\begin{equation}
	\mathcal{V} = \left\{ \theta^* : \ \theta \in \mathcal{P} \text{ s.t. } C(\theta) = 0 \right\} \ .
\end{equation}
Since $C(\theta)$ is both vanishing and minimal in $\mathcal{V}$, the same set can be defined through the optimisation statement
\begin{equation}
	\mathcal{V} = \left\{ \theta^* : \ \theta \in \mathcal{P} \text{ s.t. } \theta^* =  \arg \min C(\theta) \right\} \ .
\end{equation}
Therefore, the task of finding the points in the parameter space that satisfy constraints is the same as finding the points that minimise $C(\OB)$. When faced with multiple constraints on multiple observables or constrained quantities, $\OB_i$, we can then combine them into a single \emph{loss function}, $L$, which we wish to minimise
\begin{equation}
	L(\theta) = \sum_{i=1}^{N_C} C(\OB_i(\theta)) ,
	\label{eq:loss}
\end{equation}
where the sum runs over all the $N_C$ constraints discussed in~\cref{s:constraints}, and $L\geq0 \ \forall_{\theta}$ with $L=0$ if and only if all constraints are met. We notice that the quantity $\OB_i$ does not need to be observable, such as the $\mu_{if}$ signal strengths measured by the LHC. For example, the theoretical constraints related to BFB and unitary conditions are included as $\OB_i$ with the respective required bounds. The ability to mix measurements, limits, and theoretical constraints in the same loss function is a key strength of this methodology. Although $\OB_i$ included observables and other constrained quantities, we will often abuse terminology and refer to all $\OB_i$ as observables and the space they span as observable space, $\OB$.

\subsection{The Search Algorithm: Covariant Matrix Approximation Evolutionary Strategy}

Having framed parameter space sampling as a black-box optimisation problem, we need to choose which AI black-box optimisation algorithm to perform the task. In~\cite{deSouza:2022uhk}, three different algorithms were considered: a Bayesian optimisation algorithm, a genetic algorithm, and an evolutionary strategy algorithm. Each realised different balances of the so-called exploration (how much of the parameter space explored) vs. exploitation (how fast it converged to $\mathcal{V}$) trade-off. In this work, we will use the evolutionary strategy algorithm, which provides the fastest convergence speed.

The evolutionary strategy algorithm in question is the Covariant Matrix Approximation Evolutionary Strategy (CMAES)~\cite{hansen2006cma,Hansen:2016eid}. Evolutionary Strategies (ES) are powerful numerical optimisation algorithms from the broader field of Evolutionary Computing (EC). EC algorithms are characterised by an iterative process in which candidate solutions for a problem are tested and the best ones are used to generate new solutions. In our case, the candidate solutions are points in the parameter space, and their suitability (i.e. their \emph{fitness}) is measured by the loss function function~\cref{eq:loss}. As opposed to Genetic Algorithms, ES do not make use of genes to generate new candidate points. Instead, in ES, new candidates are \emph{sampled} from a distribution, the parameters of which are set by the best candidates from previous iterations, called generations.

In CMAES, the distribution is a highly localised multivariate normal. This normal distribution is initialised with the centre at a random point in the parameter space, and its covariant matrix is set to the identity multiplied to an overall scaling constant $\sigma$. A generation of $\lambda$ candidates is sampled from the multivariate normal and their fitness is evaluated by~\cref{eq:loss}. Next, the $\lambda$ candidates are sorted from best to worst, and the $\mu$ best candidates are used to compute a new mean of the normal distribution, as well as to approximate its covariant matrix. Intuitively, the change in mean progresses the algorithm in the direction of steepest descent of the loss function, just like a first-order optimisation method would, and the covariant matrix approximates the (local) Hessian of the loss function, just like a second-order optimisation method would. The difference, however, is that CMAES \emph{does not} compute derivatives of the loss function, and therefore it is suitable for nonconvex, ill-conditioned, and even non-continuous loss functions. This feature makes CMAES converge very quickly on a wide variety of optimisation problems. We warn, however, that the intuitive description of CMAES presented above hides many of its inner working details, which are not relevant for the study at hand, and we point the interested reader to the original references provided.

\subsection{The Novelty Reward: Histogram Based Outlier System}

In~\cite{deSouza:2022uhk} CMAES was shown to converge very fast compared to other AI black-box search algorithms. However, it was also observed that CMAES has limited exploration capacity due to the highly localised nature of the multivariate normal from which new candidate solutions are drawn. To mitigate this, in~\cite{deSouza:2022uhk} many independent runs of CMAES with restarting heuristics were performed to draw a more global picture of the valid region of the parameter space. In this work, we present a novel approach to promote exploration by adding a \emph{novelty reward} to the loss function. To achieve this, we will compute the density of valid points already found and add it to the loss function as penalty. In this way, the loss function will be minimal not only when the constraints are satisfied, but also away valid regions that have already been found, which pushes CMAES to explore new regions and effectively working as a \emph{novelty reward}.

The addition of a penalty to the loss function in~\cref{eq:loss} might produce new local minima. For example, consider the addition of various penalties, $p_j$, each taking values $p_j\in[0,1]$. This will create a competition between penalties $p_j$, and observable optimisation $C(\OB_i)$, when $\sum_i C(\OB_i)\simeq \sum_j p_j$, producing local minima away from $\sum_i C(\OB_i)=0$ and therefore spoiling our attempt to find good points. In order to prevent this, we artificially raise the value of the loss function outside the valid region by one so that such competition never arises, i.e. we consider a new version of $L$, $\tilde L$,
\begin{equation}
	\tilde L(\theta) =\begin{cases}
		1 +L(\theta) & \text{if } L(\theta) > 0 \\
		0            & \text{if } L(\theta) = 0
	\end{cases} \ ,
\end{equation}
which guarantees that the total loss function, $L_T$, including $N_p$ penalties
\begin{equation}
	L_T(\theta) = \tilde L(\theta) + \frac{1}{N_p}\sum_{i=1}^{N_p} p_i \ ,
	\label{eq:loss-new}
\end{equation}
is still positive semi-definite, and such that for a valid point we have $ \sum_i C(\OB_i) = 0 \Rightarrow 0 \leq L_T \leq 1$ in a way that the density penalty does not compete with the constraints since, i.e. for invalid points we always have $L_T>1$.

Having defined how penalties can be added to the loss function without spoiling our implementation of a black-box optimisation algorithm to find valid points, we now have to choose how to compute and quantify the penalty to produce the \emph{novelty reward}. The first thing we need to address is how to compute the point density. This task is known in the Machine Learning (ML) literature as \emph{density estimation}, and for large multidimensional datasets it is a very challenging task. Furthermore, not only we want to be able to estimate the point density accurately, we do not want the density estimation to be prohibitively slow to compute. After some preliminary exploratory experimentation, we identified the Histogram Based Outlier System (HBOS)~\cite{Goldstein2012HistogrambasedOS} as a fast and easy to implement solution.\footnote{Other possibilities were explored, such as One-Class Support Vector Machines, Isolation Forest, Kernel Density Estimation, among others, but with considerable computational cost increase. A systematic study of alternative novelty reward models is left for future work.} HBOS has also been previously explored in the context of model-independent searches for new physics using anomaly detection~\cite{CrispimRomao:2020ucc}. HBOS works by fitting a histogram with a fixed number of bins, $N_{bins}$, to each dimension, i.e for each parameter. A density penalty for a point $\theta$ is obtained by summing the heights, $h_j$, of the bins on which the values of the parameters, $\theta_j$, belong.\footnote{The usage of histograms suggests that HBOS suffers from the so-called \emph{curse of dimensionality}. In our studies below, we will see that this manifests non-trivially as an interplay between CMAES exploration and the geometry and topology of the valid region of the parameter space.} The penalties are normalised to be $p \in [0,1]$, so that a novelty point has penalty $0$ and a point too similar to the ones already seen has maximal penalty $1$. Furthermore, we notice that the penalty over the parameter space density needs not to be over all parameters, $\theta=(\alpha_1, \beta_2, \dots)$, but can be \emph{focused} on only a subset of these, $\{\theta_j\}$ -- this is especially useful to promote focused exploration in parameters of interest.

Whilst the discussion above focusses on parameter space density penalty, it can be extended to other spaces. Of particular interest, which we will include in our study, is the space of physical quantities, $\OB$. This will allow us to explore not only novel areas of the parameter space, but -- perhaps more importantly -- cover novel areas of the observables space, i.e. to explore novel phenomenological aspects of the model. To do so, we need to train a penalty $p$ not on the values of the parameters, $\theta$, but on its resulting physical quantities, $\{\OB_i\}$, where the set $\{\OB_i\}$ can be of all or a subset of $\OB_i$.

In our scans, we will include the novelty reward in both the parameter space and in the observable space, and in each case we will study penalties focused subsets of the parameters and the observables.

\subsection{Further Implementation Details}

We now discuss some implementation details of the ideas presented above. We implemented CMAES using \verb|DEAP| - Distributed Evolutionary Algorithms in Python~\cite{fortin2012deap}. The function $C$ was slightly modified so as to force all values of $C(\OB_i)$ to be nominally similar. To achieve this, we have implemented in the code $C(\OB_i) \to \log(C(\OB_i)+1)$ which retains the same properties: positive semidefiniteness, continuous, and monotonically increasing away from the allowed interval. Furthermore, to prevent any constraint $C(\OB_i)$ from dominating over any other, we rescale them at each generation to be bounded by $[0,1]$ using \verb|scikit-learn|~\cite{pedregosa2011scikit} \verb|MinMaxScaler| before computing $\tilde L$ and the final loss~\cref{eq:loss-new}. We used \verb|pyod|~\cite{zhao2019pyod} implementation of the HBOS~\cite{Goldstein2012HistogrambasedOS}, and set $N_{bins}=100$, observing a considerable computational overhead for higher values.\footnote{The Evolutionary Strategy with Novelty Reward implementation is made available in\url{https://gitlab.com/miguel.romao/evolutionary-strategy-novelty-detection-3hdm}.}

The constraints that were checked in our main computational loop are listed in~\cref{s:appendix_A}. For collider limits on novel scalars, we used \verb|HiggsTools| version \verb|1.1.3|~\cite{Bahl:2022igd}. As we perform the signal strength, $\mu_{ij}$, checks in our main computational loop, we only use the \verb|HiggsBounds| functionality of \verb|HiggsTools|, implemented using the \verb|HiggsBounds| version \verb|5| input files using the Python script provided by the \verb|HiggsTools| authors, and used the \verb|HiggsBounds| dataset version \verb|1.4|.

For each scan, we ran a total of $100$ independent runs, each with a maximum of $2000$ generations. We use the CMAES default parameters, which set the population size, $\lambda$, and the number of best candidates, $\mu$, using a heuristic. The values for our problem were automatically set to $\lambda=12$ and $\mu=6$, which means that each scan will at most evaluate $2000\times12\times100=2.4 \times 10^6$ points. CMAES has very few parameters to be defined by the user, only the initial mean of the multivariate normal and the overall scale of the covariance $\sigma$. For each run, we set the mean to a random point in the parameter space and initialise CMAES with $\sigma = 1$.

Furthermore, we follow the methodology in~\cite{deSouza:2022uhk} where we define all operations related to CMAES and density estimation not over the parameter space, but over a hypercube $[0,1]^{d_\mathcal{P}}$, where $d_\mathcal{P}$ is the number of parameters, that maps to the parameter space $\mathcal{P}$. This allows us to treat all parameters in an equal nominal footing, avoiding any potential pathologies arising from having different parameters spanning many different orders of magnitude. {For all scans the parameter space bounds are the same used for the random scans, defined in~\cref{eq:scanparameters}.}

The final loss to be optimised to explore the parameter space is
\begin{equation}
	L_T(\theta) = \tilde L(\theta) + \frac{1}{2} (p_\mathcal{P}(\{\theta_j\})+p_\OB(\{\OB_j(\theta)\})) ,
\end{equation}
where $p_\mathcal{P}(\{\theta_j\})$ is the density penalty computed
over the subset of parameters $\{\theta_j\}$ effectively working as a
novelty reward in $\mathcal{P}$, $p_\OB(\{\OB_j(\theta)\})$ is the
density penalty computed  over the subset of observables $\{\OB_j\}$
effectively working as a novelty reward in $\mathcal{O}$, $N_C$ is the
number of constraints. As discussed previously, we present different
scans for different choices of $\{\theta_j\}$ and $\{\OB_j\}$ to be
included in the computation of $p_\mathcal{P}$ and $p_\OB$ to promote
\emph{focused} scans.


\section{Analysis and results}
\label{s:results}

We now present and analyse the results for multiple scans using the ideas presented in~\cref{s:AI}, and two random sampling scan strategies: purely random over the whole parameter space and $50\%$ away from the alignment limit defined by~\cref{Al-2}. We present two analysis, one where the \verb|HiggsBounds| constraints using \verb|HiggsTools| were not included in the loss function, and one where it has. The scan without \verb|HiggsTools| runs faster in both computational overhead and CMAES convergence (to be discussed below), which allowed us to experiment with our methodology. The impact of
	{using \texttt{HiggsTools}, \textit{a posteriori}}, on points obtained without including it in the loop is studied. We then perform a second analysis where we included \verb|HiggsTools| in the loop to show how one can include multiple constraints from different sources and still be able to explore the whole parameter space of the model.

We start with scans that do not take into account \verb|HiggsTools| results in the loss function. All scans are performed over the whole $16$-dimensional parameter space and all have the same $61$ constraints. We then include \verb|HiggsTools| in the optimisation loop by adding the respective contribution to the loss function. All scans and their details can be seen in~\cref{tab:scans}. As will be discussed in~\cref{sec:convergence_metrics}, \verb|HiggsTools| reduces the number of successful runs by a factor of 2, and therefore for these runs we allow for 200 instead of 100 scans.
\begin{table}[htb]
	\makebox[1.0\textwidth]{
		\begin{tabular}{p{2.8cm}p{4.75cm}|p{0.25cm}p{0.95cm}p{0.95cm}p{2.75cm}p{2.75cm}}
			\hline\hline
			Sampling                                & Scan                                                  &  & $d_\mathcal{P}$ & $N_{C}$ & $p_{\mathcal{P}}$                            & $p_{\mathcal{O}}$                               \\ \hline
			\multirow{2}{*}{Random}                 & Completely random                                     &  & $16$            & $61$    & N/A                                          & N/A                                             \\
			                                        & Alignment limit: AL-2                                 &  & $16$            & $61$    & N/A                                          & N/A                                             \\ \hline
			\multirow{6}{*}{CMAES}                  & No penalty                                            &  & $16$            & $61$    & None                                         & None                                            \\
			                                        & Parameter novelty reward                              &  & $16$            & $61$    & $16$ parameters                              & None                                            \\
			                                        & Observable novelty reward                             &  & $16$            & $61$    & None                                         & $61$ quantities $\OB_i$                         \\
			                                        & $\alpha_1$, $\beta_1$, $\alpha_2$, $\beta_2$ focus    &  & $16$            & $61$    & $\alpha_1$, $\beta_1$, $\alpha_2$, $\beta_2$ & None                                            \\
			                                        & $\mu_{ggF, \gamma\gamma}$, $\mu_{ggF, Z\gamma}$ focus &  & $16$            & $61$    & None                                         & $\mu_{ggF, \gamma\gamma}$, $\mu_{ggF, Z\gamma}$ \\
			                                        & $m_{H_1^+}$, $m_{H_2^+}$ focus                        &  & $16$            & $61$    & $m_{H_1^+}$, $m_{H_2^+}$                     & None                                            \\ \hline
			\multirow{5}{*}{CMAES}                  & Parameter novelty reward                              &  & $16$            & $67$    & $17$ parameters                              & None                                            \\
			\multirow{5}{*}{w/ \texttt{HiggsTools}} & Observable novelty reward                             &  & $16$            & $67$    & None                                         & $67$ quantities $\OB_i$                         \\
			                                        & $\alpha_1$, $\beta_1$, $\alpha_2$, $\beta_2$ focus    &  & $16$            & $67$    & $\alpha_1$, $\beta_1$, $\alpha_2$, $\beta_2$ & None                                            \\
			                                        & $\mu_{ggF, \gamma\gamma}$, $\mu_{ggF, Z\gamma}$ focus &  & $16$            & $67$    & None                                         & $\mu_{ggF, \gamma\gamma}$, $\mu_{ggF, Z\gamma}$ \\
			                                        & $m_{H_1^+}$, $m_{H_2^+}$ focus                        &  & $16$            & $67$    & $m_{H_1^+}$, $m_{H_2^+}$                     & None                                            \\ \hline\hline
		\end{tabular}
	}
	\caption{\label{tab:scans}List of scans performed, where $d_{\mathcal{P}}$ is the number of parameters, $N_C$ the number of constraints, $p_{\mathcal{P}}$ the novelty reward over the parameter space, $p_{\mathcal{O}}$ the novelty reward over the space of physical quantities $\OB_i$.}
\end{table}

\subsection{Rewarding Exploration in the Parameter Space\label{subsec:parameter_space_reward}}

\begin{figure}[H]
	\makebox[\textwidth][c]{
		\begin{subfigure}[t]{0.375\textwidth}
			\includegraphics[width=\linewidth]{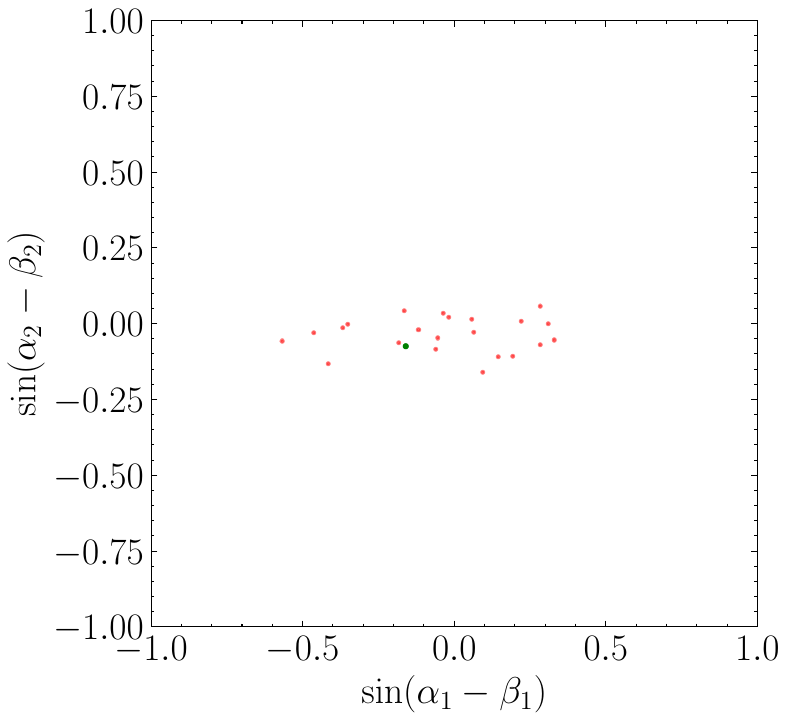}
			\caption{Random sampling}
			\label{fig:sa1b1_vs_sa2b2_no_HT_k9}
		\end{subfigure}
		\begin{subfigure}[t]{0.375\textwidth}
			\includegraphics[width=\linewidth]{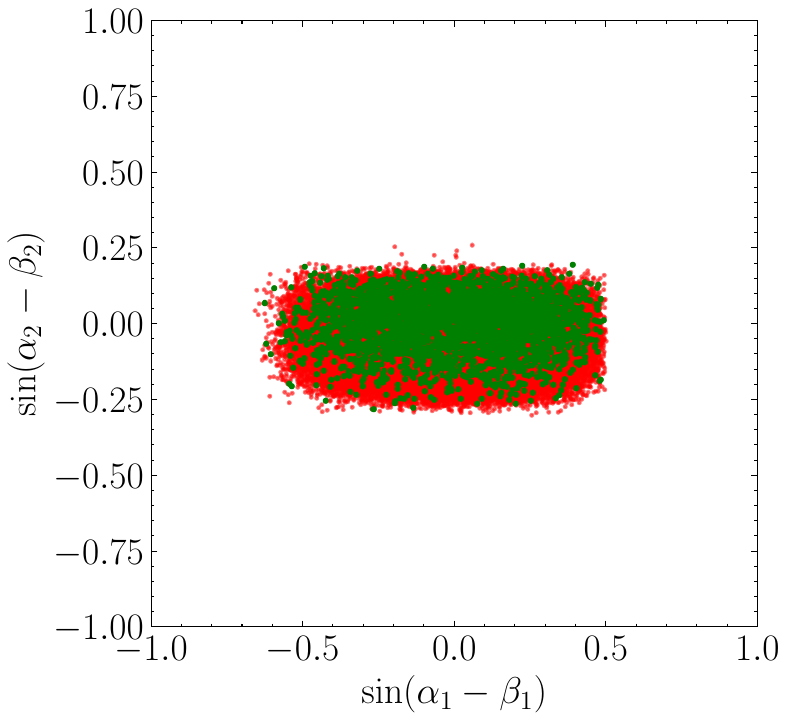}
			\caption{Alignment limit: AL-2}
			\label{fig:sa1b1_vs_sa2b2_no_HT_k16}
		\end{subfigure}
		\begin{subfigure}[t]{0.375\textwidth}
			\includegraphics[width=\linewidth]{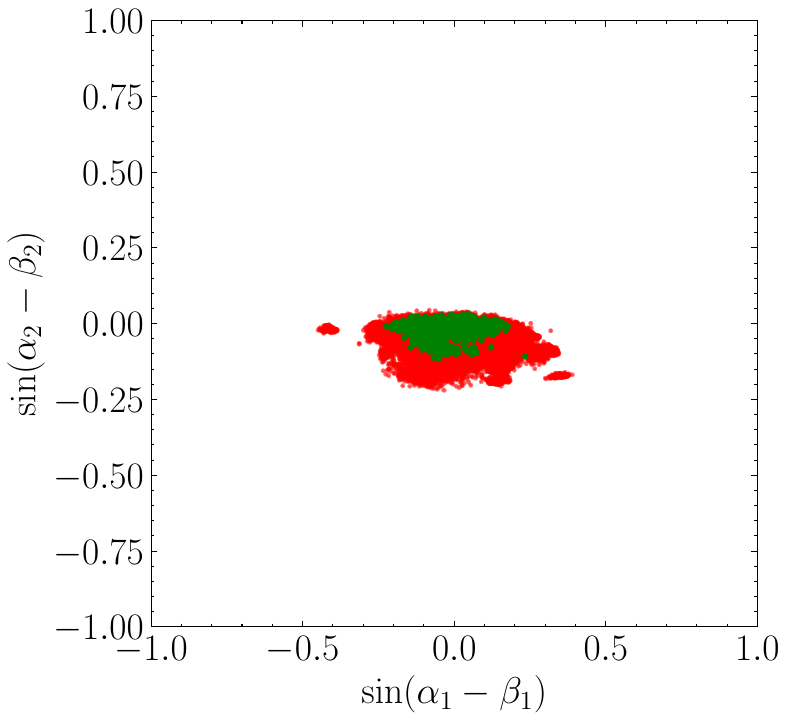}
			\caption{CMAES no penalty}
			\label{fig:sa1b1_vs_sa2b2_no_HT_cmaes_vanilla}
		\end{subfigure}
	}\\
	\makebox[\textwidth][c]{
		\begin{subfigure}[t]{0.375\textwidth}
			\includegraphics[width=\linewidth]{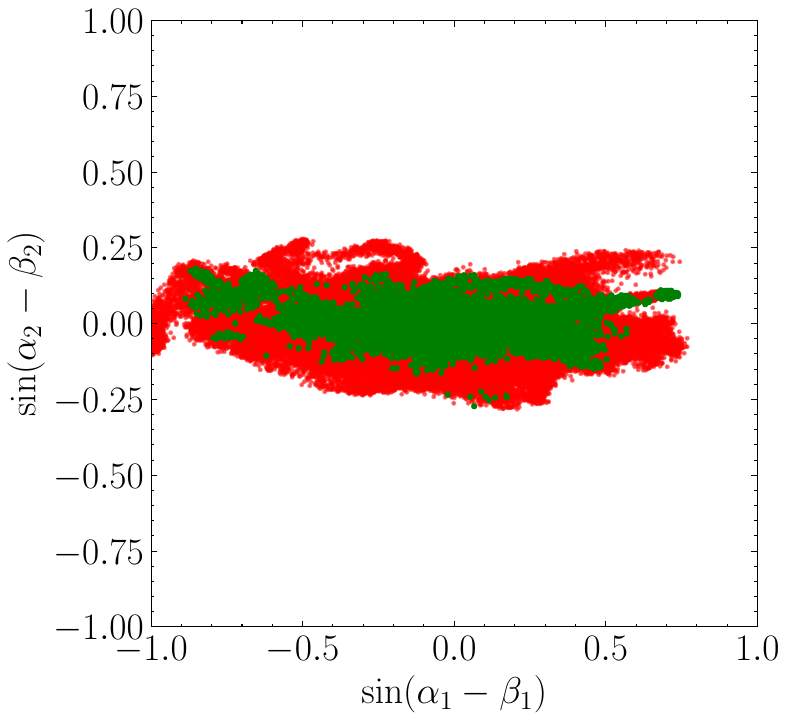}
			\caption{CMAES with parameter novelty reward}
			\label{fig:sa1b1_vs_sa2b2_no_HT_cmaes_parameter_penalty}
		\end{subfigure}
		\begin{subfigure}[t]{0.375\textwidth}
			\includegraphics[width=\linewidth]{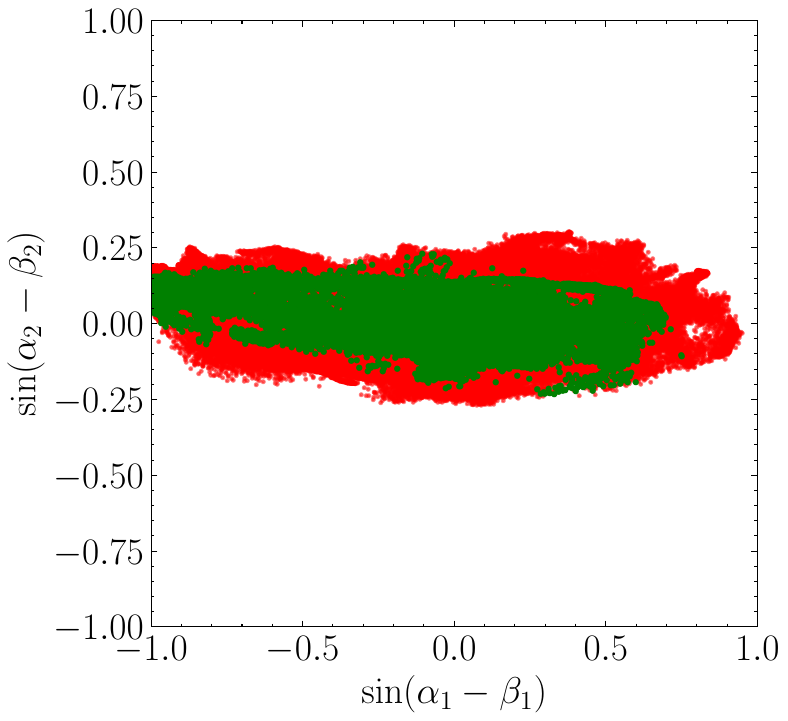}
			\caption{CMAES with $\alpha_1, \beta_1, \alpha_2, \beta_2$ focus}
			\label{fig:sa1b1_vs_sa2b2_no_HT_cmaes_focused_a1_b1_a2_b2}
		\end{subfigure}
	}
	\caption{$(\sin(\alpha_1 - \beta_1), \sin(\alpha_2-\beta_2))$ scatter plot for different runs without novelty reward. A point surviving \texttt{HiggsTools} is represented in green, otherwise in red.}
	\label{fig:sa1b1_vs_sa2b2_no_HT}
\end{figure}

We first study the implementation of CMAES with and without parameter space reward to show the enhanced exploration capabilities of CMAES when including a parameter density penalty in the loss function. In~\cref{fig:sa1b1_vs_sa2b2_no_HT} we show the scatter plot of the $(\sin(\alpha_1 - \beta_1), \sin(\alpha_2-\beta_2))$ plane for different runs. In particular, we exhibit the difficulty of random sampling finding valid points, with \cref{fig:sa1b1_vs_sa2b2_no_HT_k9} showing only 23 valid points, of which only one passed the \verb|HiggsBounds| constraints. These were obtained from a scan that sampled an estimated $\mathcal{O}(10^{13})$.\footnote{We can only estimate the number of points as it would have been prohibitive to store all non-valid points. Therefore, we measured how long the scan took to process a few thousand points, and kept a loose track of the random sampling run to produce the estimate.} In~\cref{fig:sa1b1_vs_sa2b2_no_HT_k16} we show the points obtained by sampling within 50\% of the alignment limit, where the allowed points are highly constrained with $\alpha_1 \simeq \beta_1$, an expected result due to the highly constraining bounds from collider measurements of the Standard Model-like Higgs boson decay channels. In the same graph, we can also observe the boundaries imposed by the alignment limit, as $|\sin(\alpha_1 - \beta_1) | \simeq 0.5$ {and how $\alpha_2$ are $\beta_2$ are considerably more constrained leading to $|\sin(\alpha_2 - \beta_2) | \simeq 0.25$}. In the next plot,~\cref{fig:sa1b1_vs_sa2b2_no_HT_cmaes_vanilla}, we show the result of the CMAES scan without further exploration. We observe a funnelling of the results into a single region $\alpha_1,\ \beta_1,\ \alpha_2,\ \beta_2 \simeq 0$, clearly providing little coverage of the parameter space, although still providing far more valid points than the random sampler. This lack of exploration of CMAES was first observed in~\cite{deSouza:2022uhk} and is easily understood from an algorithm point of view as CMAES does not have built-in mechanisms to escape a minimum (global or minimal).\footnote{In~\cite{deSouza:2022uhk} CMAES was endowed with a restart strategy, which mitigates this and allowed CMAES to draw a more complete picture of the valid regions of the parameter space. In this paper, we have not implemented this, as our focus is on developing a novelty reward driven exploration.}
Despite the seemingly lacklustre result, the points in~\cref{fig:sa1b1_vs_sa2b2_no_HT_cmaes_vanilla} obtained by CMAES were obtained in a quick run of only around $\mathcal{O}(10^3)$ attempts, providing around ten orders of magnitude sampling efficiency improvement over the random scan (see~\cref{sec:convergence_metrics} for a more detailed discussion on convergence metrics and performance). This allows us to change the sampling logic as to explore the parameter space once the sampler converges into a valid subregion of the parameter space. As explained in the preceding sections, this is achieved by including a parameter density penalty. In~\cref{fig:sa1b1_vs_sa2b2_no_HT_cmaes_parameter_penalty} we show the result for the CMAES runs when activating the novelty reward for all parameters, i.e. using a density penalty for all parameters. We can immediately see a much larger region of the parameter space scanned, especially beyond the alignment limit in the $\sin(\alpha_1-\beta_1)$ direction. We can also observe the first \emph{artefact} of this methodology: we can see sequences of points, akin to the trail of a paintbrush, in this plane. These trails are in fact the path that CMAES has covered while exploring the parameter space away from previously found points. By introducing the parameter penalty over all parameter space, we were able to find novel points away from the alignment limit. However, because the penalty is computed over all the parameter space, CMAES has no incentive to explore \emph{interesting} regions of the parameter space, as it can reduce the density penalty by spreading across parameters which have little impact on the constraints.\footnote{This can be seen as a variation of the \emph{curse of dimensionality}.} To mitigate this, we \emph{focus} the parameter density penalty on the four parameters described by these scatter plots, $\alpha_1, \beta_1, \alpha_2, \beta_2$. The resulting points can be seen in~\cref{fig:sa1b1_vs_sa2b2_no_HT_cmaes_focused_a1_b1_a2_b2}, where we see that CMAES was able to spread its exploration even more in the $(\sin(\alpha_1 - \beta_1), \sin(\alpha_2-\beta_2))$ plane. More interestingly, the points that subsequently pass \verb|HiggsBounds|, shown in green, cover a much larger region than those obtained using the sampling around the alignment limit, although the latter has arguably a better coverage over $\sin(\alpha_2-\beta_2)$ values more uniformly over different values of $\sin(\alpha_1-\beta_1)$.

\begin{figure}[H]
	\makebox[\textwidth][c]{
		\begin{subfigure}[t]{0.375\textwidth}
			\includegraphics[width=\linewidth]{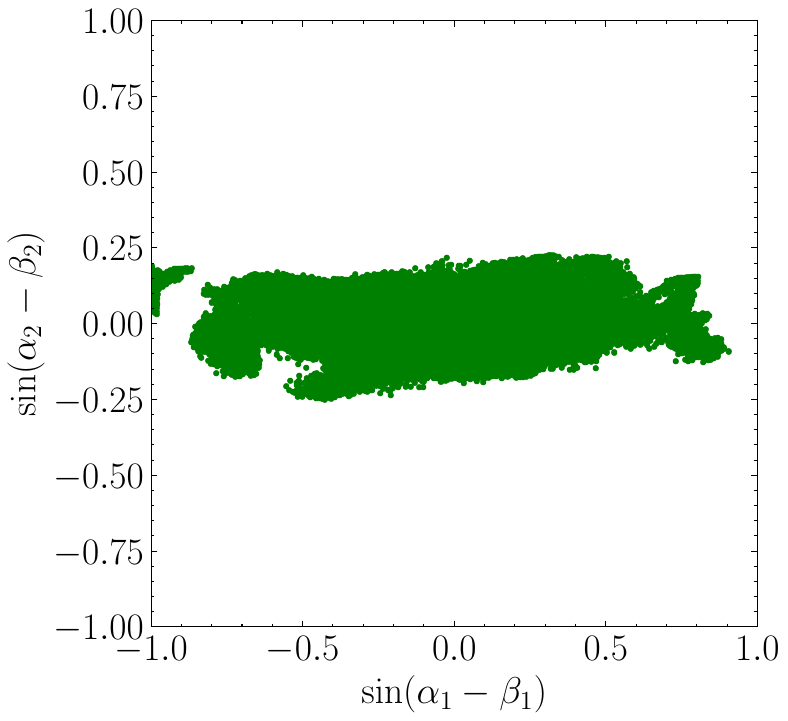}
			\caption{CMAES with parameter novelty reward}
			\label{fig:sa1b1_vs_sa2b2_w_HT_cmaes_parameter_penalty}
		\end{subfigure}
		\begin{subfigure}[t]{0.375\textwidth}
			\includegraphics[width=\linewidth]{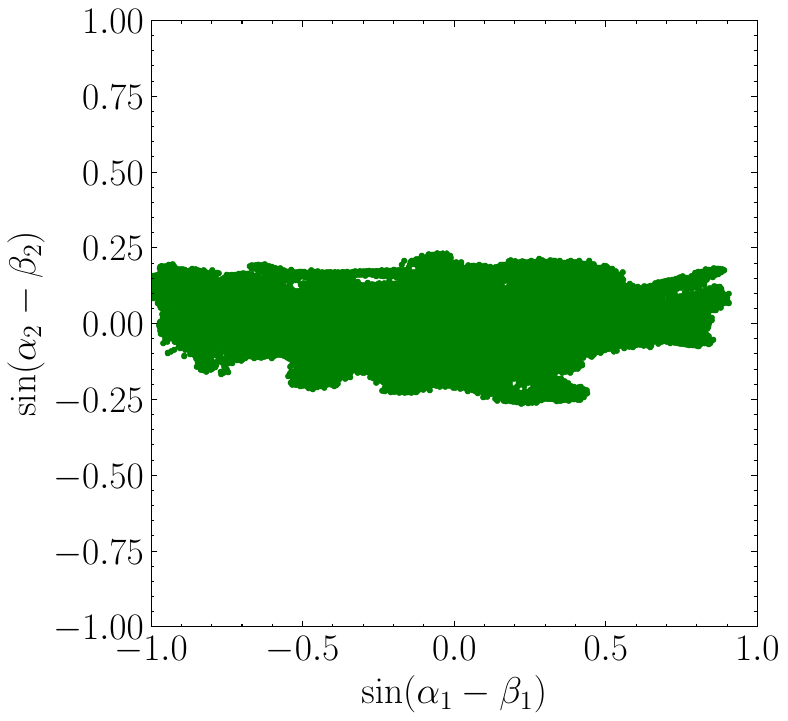}
			\caption{CMAES with $\alpha_1, \beta_1, \alpha_2, \beta_2$ focus}
			\label{fig:sa1b1_vs_sa2b2_w_HT_cmaes_focused_a1_b1_a2_b2}
		\end{subfigure}
	}
	\caption{$(\sin(\alpha_1 - \beta_1), \sin(\alpha_2-\beta_2))$ scatter plot for different runs with novelty reward and \texttt{HiggsTools} constraints included in the loss function.}
	\label{fig:sa1b1_vs_sa2b2_w_HT}
\end{figure}

The above scans were produced by performing a run without checking for the constraints coming from \verb|HiggsBounds| provided by \verb|HiggsTools|. The survival rate against \verb|HiggsTools| of the points obtained using these scans is $3\%-5\%$, or, in other words, a factor of $1/20$ or less. Furthermore, the execution time with \verb|HiggsTools| increases by a factor of around $3$. To first approximation, checking for \verb|HiggsBounds| in the loop can slow down the process of finding good points by an expected factor of $60$ or more. On the other hand, as we can see in~\cref{fig:sa1b1_vs_sa2b2_no_HT} not using \verb|HiggsTools| leads to too many invalid points, and depriving CMAES of this information will only prevent it from finding points that survive \verb|HiggsBounds|. In~\cref{fig:sa1b1_vs_sa2b2_w_HT} we present the first scans with \verb|HiggsTools| in the loop to check for \verb|HiggsBounds| constraints. The scans presented in~\cref{fig:sa1b1_vs_sa2b2_w_HT_cmaes_parameter_penalty} and~\cref{fig:sa1b1_vs_sa2b2_w_HT_cmaes_focused_a1_b1_a2_b2} are direct analogous to those presented in~\cref{fig:sa1b1_vs_sa2b2_no_HT_cmaes_parameter_penalty} and~\cref{fig:sa1b1_vs_sa2b2_no_HT_cmaes_focused_a1_b1_a2_b2}, respectively. In both cases we observe a far wider coverage of the parameter space than before, showcasing the importance of providing \verb|HiggsTools| feedback to CMAES. More importantly, both runs covered the space of valid points around the alignment limit in this plane, but went far beyond in the $\alpha_1, \beta_1$ subspace.

\begin{figure}[H]
	\makebox[\textwidth][c]{
		\begin{subfigure}[t]{0.375\textwidth}
			\includegraphics[width=\linewidth]{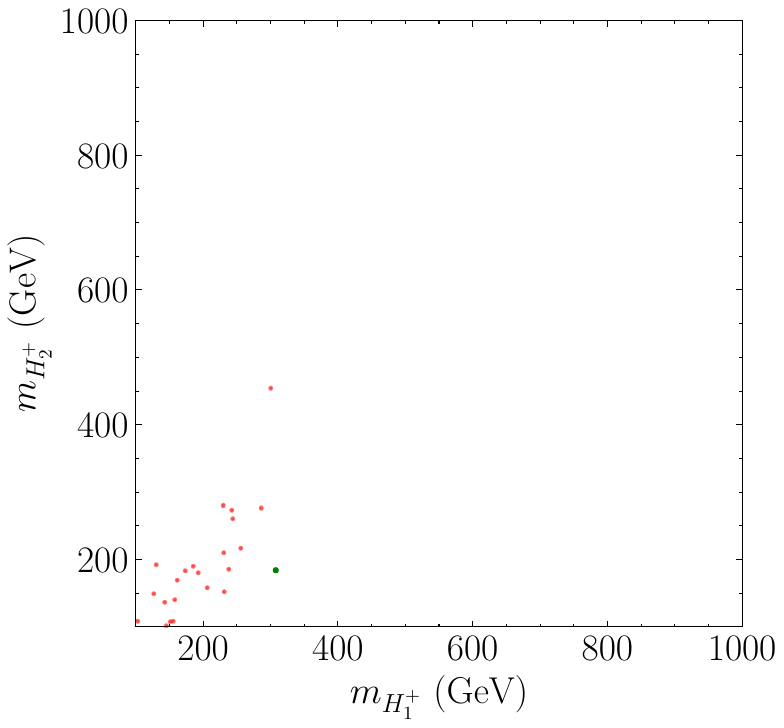}
			\caption{Random sampling}
			\label{fig:mC1_mC2_no_HT_k9}
		\end{subfigure}
		\begin{subfigure}[t]{0.375\textwidth}
			\includegraphics[width=\linewidth]{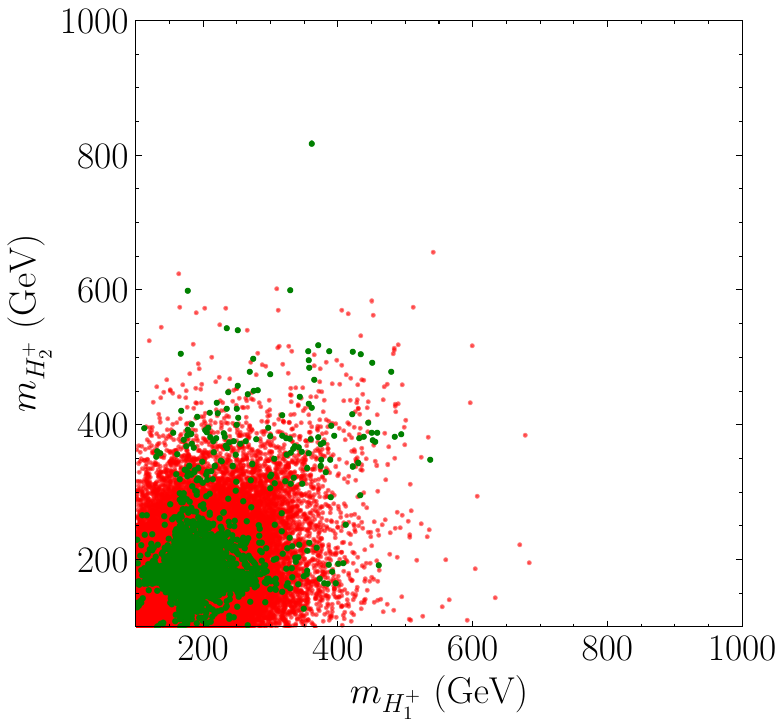}
			\caption{Alignment limit: AL-2}
			\label{fig:mC1_mC2_no_HT_k16}
		\end{subfigure}
		\begin{subfigure}[t]{0.375\textwidth}
			\includegraphics[width=\linewidth]{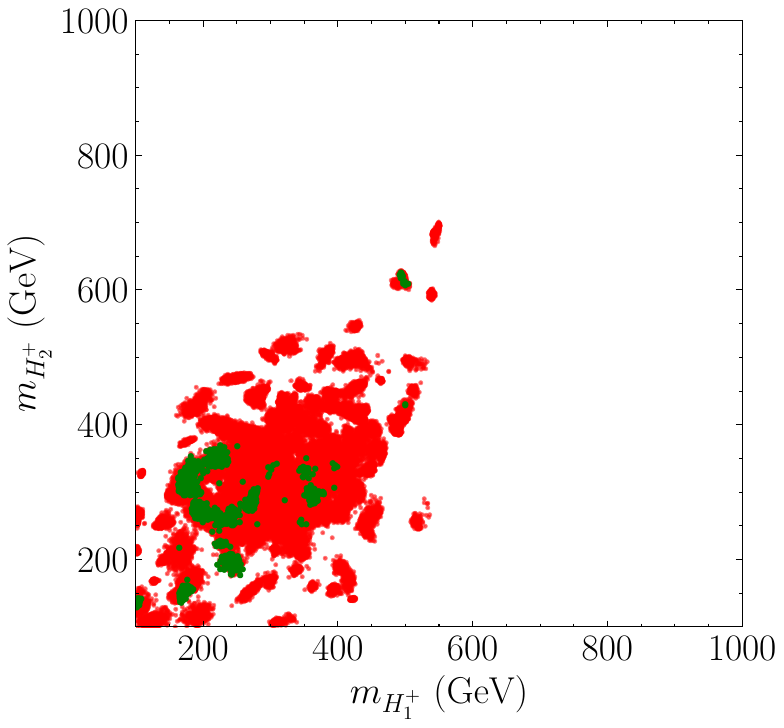}
			\caption{CMAES no penalty}
			\label{fig:mC1_mC2_no_HT_cmaes}
		\end{subfigure}
	}\\
	\makebox[\textwidth][c]{
		\begin{subfigure}[t]{0.375\textwidth}
			\includegraphics[width=\linewidth]{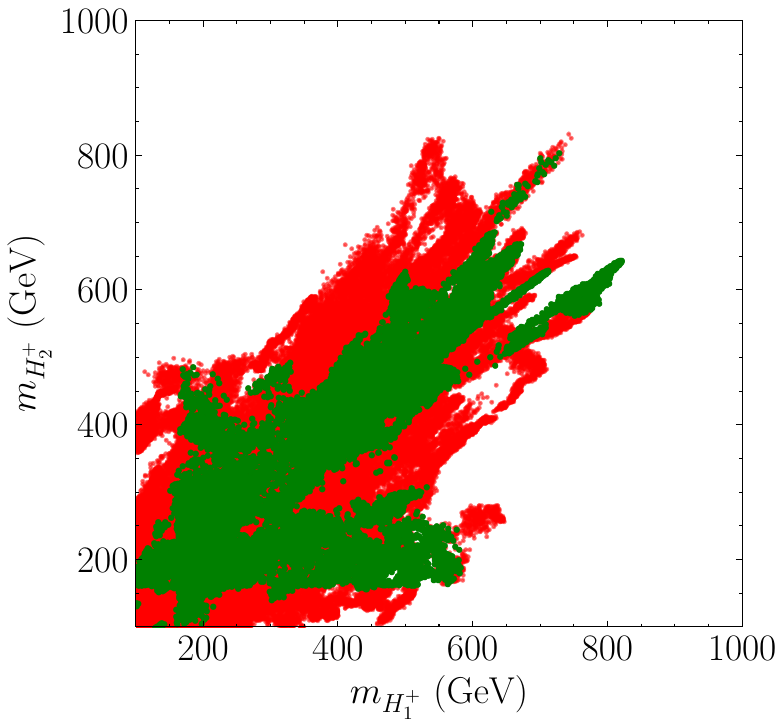}
			\caption{CMAES with parameter novelty reward}
			\label{fig:mC1_mC2_no_HT_cmaes_parameter_penalty}
		\end{subfigure}
		\begin{subfigure}[t]{0.375\textwidth}
			\includegraphics[width=\linewidth]{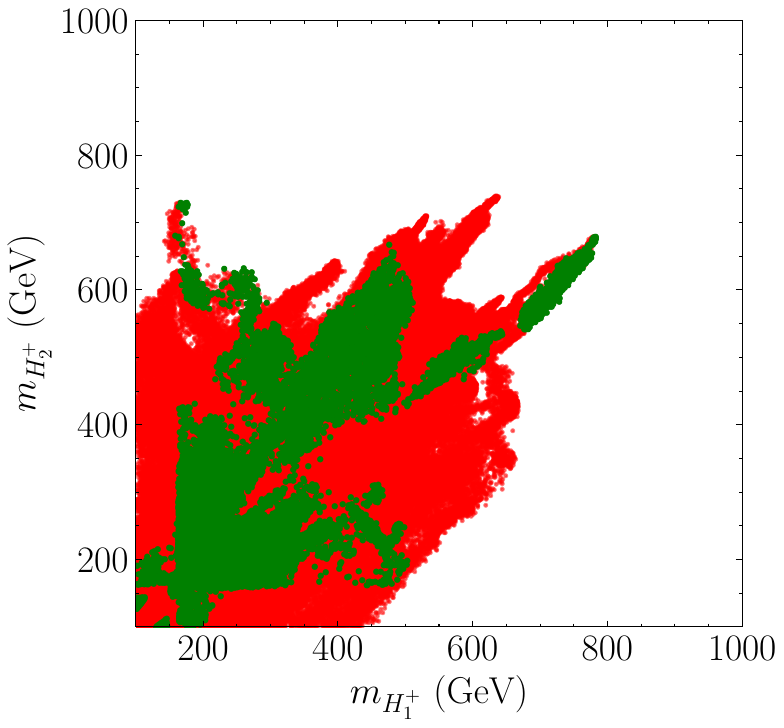}
			\caption{CMAES with $m_{H^+_1}, m_{H^+_2}$ focus}
			\label{fig:mC1_mC2_no_HT_cmaes_focused_mC1_mC2}
		\end{subfigure}
	}
	\caption{$(m_{H^+_1}, m_{H^+_2})$  scatter plot for different runs without novelty reward. A point surviving \texttt{HiggsTools} is represented in green, otherwise in red.}
	\label{fig:mC1_mC2_no_HT}
\end{figure}

We now turn to the masses of the charged scalars, which are constrained by direct searches. In~\cref{fig:mC1_mC2_no_HT} we show the points obtained in the $(m_{H^+_1},m_{H^+_2})$ plane. The logic is similar to the previous discussion, with the difference that the last plot,~\cref{fig:mC1_mC2_no_HT_cmaes_focused_mC1_mC2}, shows the points from a scan where the penalty was \emph{focused} on the charge scalar masses instead of the mixing angles. From the scan around the alignment limit,~\cref{fig:mC1_mC2_no_HT_k16}, we can observe how \verb|HiggsBounds| affects the valid region, especially for small values of scalar masses. Interestingly, in~\cref{fig:mC1_mC2_no_HT_cmaes} we see that CMAES has provided more coverage over this cross section of the parameter space than before. This can be easily interpreted: CMAES works akin to a gradient descent algorithm, but with the performance enhanced by the approximation of local second derivative. This means that CMAES \emph{rolls down} the loss function with \emph{momentum}, following the quickest path to its minimum. This preference for a quick convergence path explains why different CMAES runs will provide similar values of the most constrained parameters, as it is through them that a path needs to be found so as to minimise the loss function. This \emph{eagerness} to converge is a feature of CMAES, which is on the \emph{ exploitive} side of the \emph{exploration-exploitation} trade off, as previously also discussed~\cite{deSouza:2022uhk}.
In~\cref{fig:mC1_mC2_no_HT_cmaes_parameter_penalty} and~\cref{fig:mC1_mC2_no_HT_cmaes_focused_mC1_mC2} we show the results of the scans with the parameter density penalty over all parameters and focused on the charged masses, respectively. We observe that both were able to cover more parameter space than the alignment limit random scan, but the scan focused on the charged scalar masses was able to cover more of the $(m_{H^+_1},m_{H^+_2})$ plane, especially in the $m_{H^+_{1/2}}\gtrsim 100$ GeV limits. This result further shows that focussing the exploration reward on subsets of parameters can help uncover novel regions overlooked by traditional scans, although in this case most of the points with $m_{H^+_{1/2}}\gtrsim 100$ GeV did not survive \verb|HiggsBounds|. {Furthermore, considering that all scans presented in~\cref{fig:mC1_mC2_no_HT}  share the same parameter boundaries, c.f.~\cref{eq:scanparameters}, these results highlight how our methodology can explore regions uncovered by the low sampling efficiency of random sampler and by the simplified priors of the alignment limit.}

{In~\cref{fig:mC1_mC2_no_HT} we can also observe a shortcoming of the method which we will encounter throughout this work. We can see that different runs lead to different sets of valid points, preventing us from drawing a global picture of the allowed points. This is due to the fact that, in each run CMAES traverses a very localised path in the parameter space until either the maximum number of generations has been reached or a local minimum stopping criterion is triggered. To mitigate this, we implemented each scan as a collection of multiple runs, but the problem has not been completely resolved. In~\cref{subsubsec:mC1_mC2,subsec:seed_points} we will present refinements of the methodology that will allow us to better fill in the \emph{voids} left by these scans.}

\begin{figure}[H]
	\makebox[\textwidth][c]{
		\begin{subfigure}[t]{0.375\textwidth}
			\includegraphics[width=\linewidth]{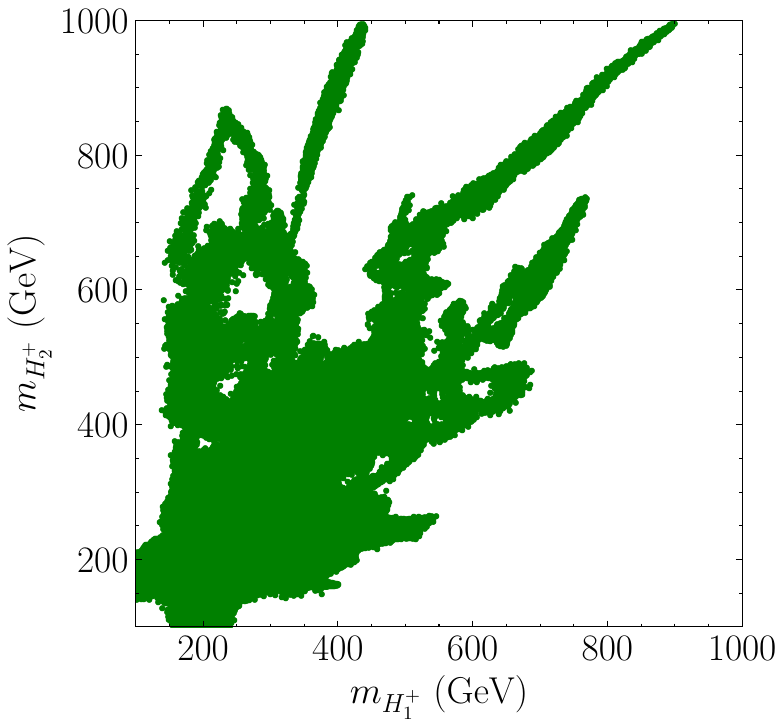}
			\caption{CMAES with parameter novelty reward}
			\label{fig:mC1_mC2_w_HT_cmaes_parameter_penalty}
		\end{subfigure}
		\begin{subfigure}[t]{0.375\textwidth}
			\includegraphics[width=\linewidth]{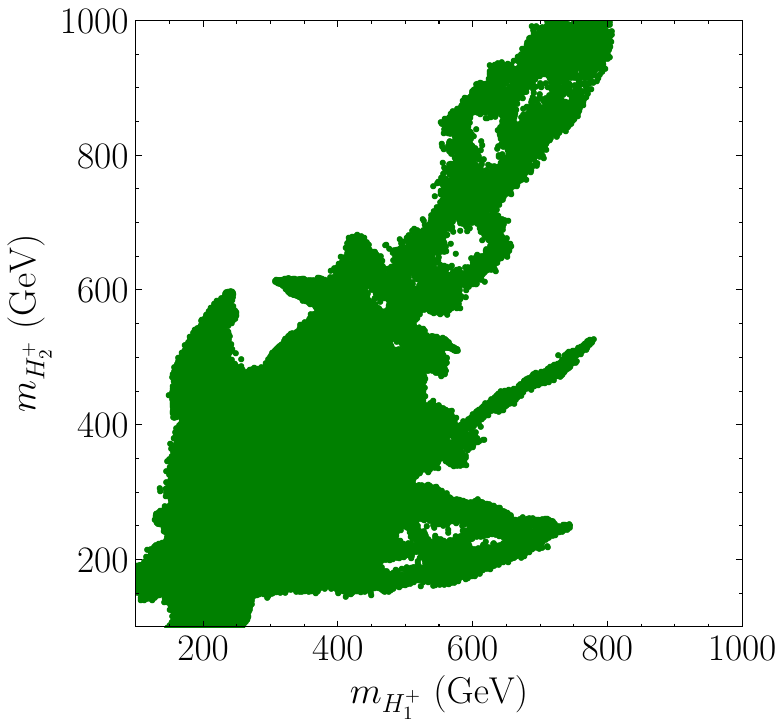}
			\caption{CMAES with $m_{H^+_1}, m_{H^+_2}$ focus}
			\label{fig:mC1_mC2_w_HT_cmaes_focused_mC1_mC2}
		\end{subfigure}
	}
	\caption{$(m_{H^+_1}, m_{H^+_2})$ scatter plot for different runs with novelty reward and \texttt{HiggsTools} constraints included in the loss function.}
	\label{fig:mC1_mC2_w_HT}
\end{figure}

Analogously to the discussion above on the mixing angles $\alpha_1,\ \beta_1,\ \alpha_2,\ \beta_2$, we now present the results with \verb|HiggsTools| included in the loop to check \verb|HiggsBounds| in~\cref{fig:mC1_mC2_w_HT}. We see that both with unfocused,~\cref{fig:mC1_mC2_w_HT_cmaes_parameter_penalty}, and charged masses focused,~\cref{fig:mC1_mC2_w_HT_cmaes_focused_mC1_mC2}, novelty reward CMAES is able to cover a much larger parameter space region than the alignment limit sampling. Furthermore, the valid points found also span a larger region than those in~\cref{fig:mC1_mC2_no_HT} that survived \verb|HiggsBounds| constraints, highlighting the importance of including \verb|HiggsTools| in the loop.

\begin{figure}[H]
	\makebox[\textwidth][c]{
		\includegraphics[width=0.475\linewidth]{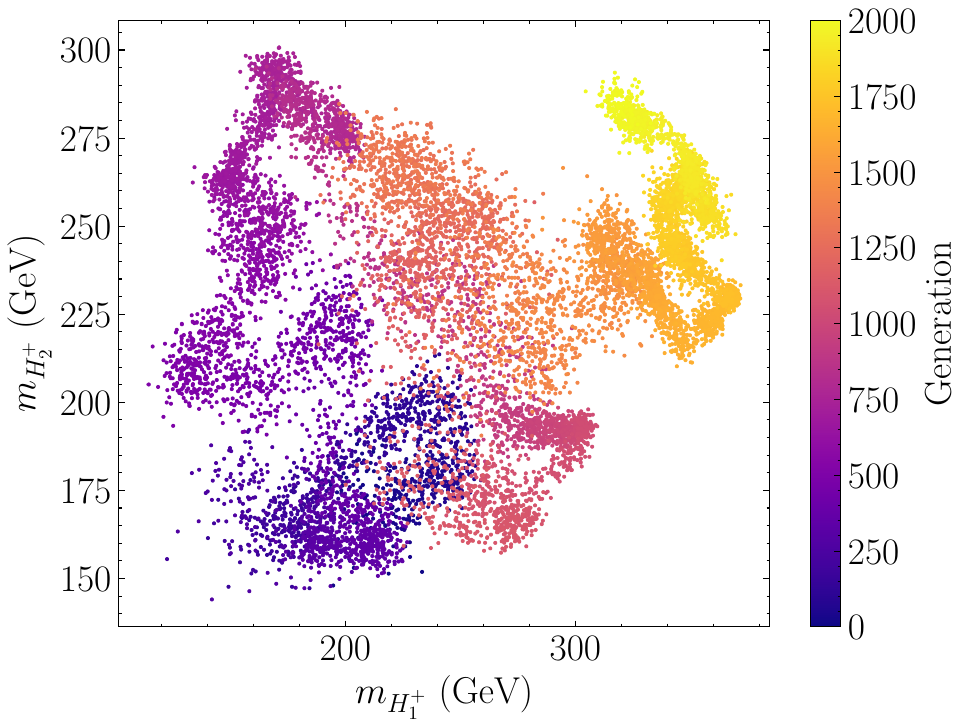}
		\includegraphics[width=0.49\linewidth]{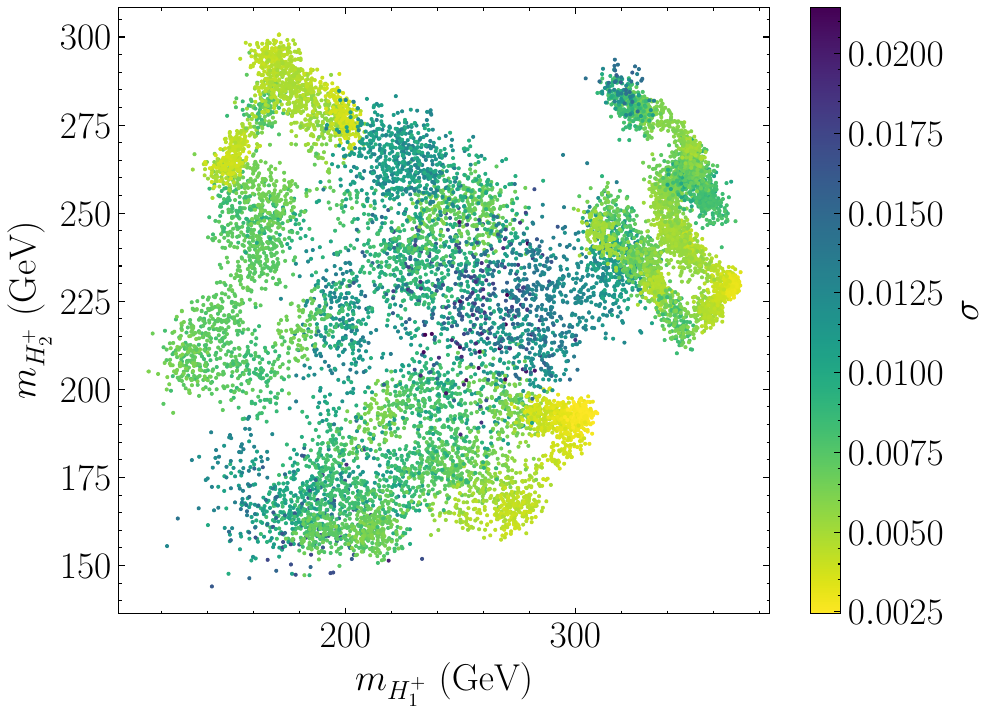}
	}
	\caption{Path of a CMAES scan with focused parameter density penalty on the $(m_{H^+_1}, m_{H^+_2})$ plane. Left: colour representing the generation number. Right: colour representing $\sigma$, the overall scale of the covariant matrix of CMAES.}
	\label{fig:mC1_mC2_run_exploration}
\end{figure}

The paths taken by CMAES while exploring the parameter space are very prominent in the scans just discussed. To better understand how CMAES is exploring, in~\cref{fig:mC1_mC2_run_exploration} we show the path traversed by a run projected onto the $(m_{H^+_1}, m_{H^+_2})$ plane. This run has converged at generation number $129$, with values $(m_{H^+_1}, m_{H^+_2})\simeq(218.6,151.4) $ GeV. At generation 129, the overall scale of the covariant matrix, given by $\sigma$ of the CMAES algorithm, is $\sigma \simeq 0.002$, a value much smaller than the initialised value of $\sigma=1$. Once converged, the density penalty is then added to the loss function forcing CMAES to explore new values of the parameters, as can be observed in the left pane. As it explores, CMAES might be slowed down by the penalty; this leads the algorithm to increase $\sigma$ to find new good points farther away. On the right pane we see this dynamical adaptation of $\sigma$ by CMAES, with higher (lower) values of $\sigma$ leading to more (less) localised samplings. This ability to adaptively increase $\sigma$ when slowed down also provides CMAES with the capacity to escape local minima, and in our case it provides a way of forcing CMAES to move away from where it has been.

\subsubsection{The $m_{H^+_{1,2}} \leq 150$ GeV Region\label{subsubsec:mC1_mC2}}

\begin{figure}[H]
	\makebox[\textwidth][c]{
		\begin{subfigure}{0.375\textwidth}
			\includegraphics[width=\linewidth]{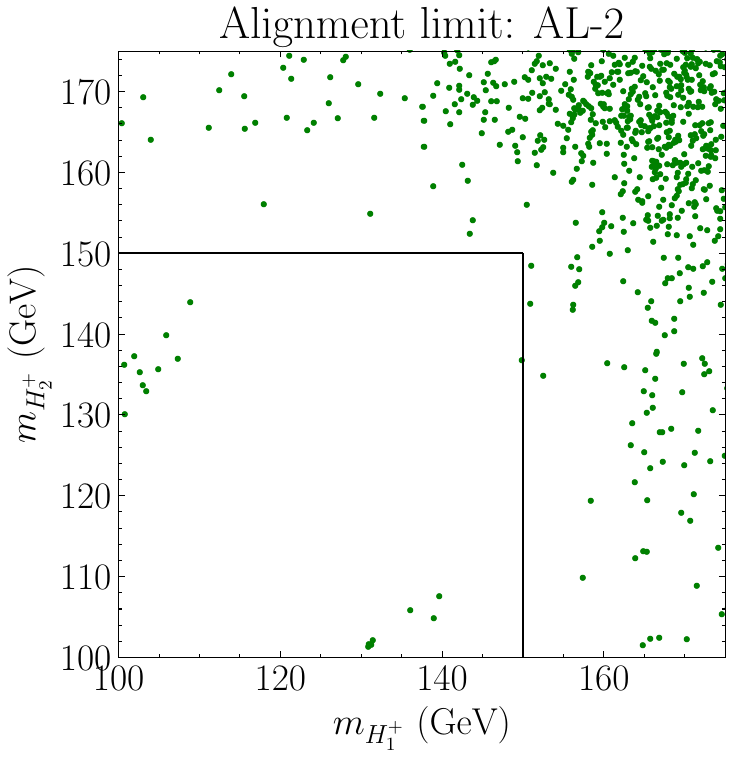}
		\end{subfigure}
		\begin{subfigure}{0.375\textwidth}
			\includegraphics[width=\linewidth]{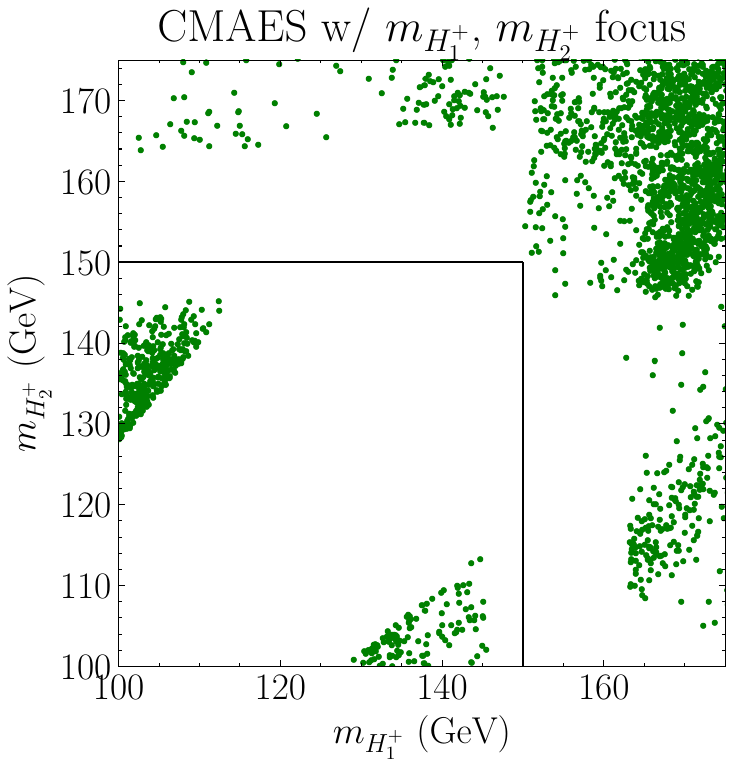}
		\end{subfigure}
		\begin{subfigure}{0.375\textwidth}
			\includegraphics[width=\linewidth]{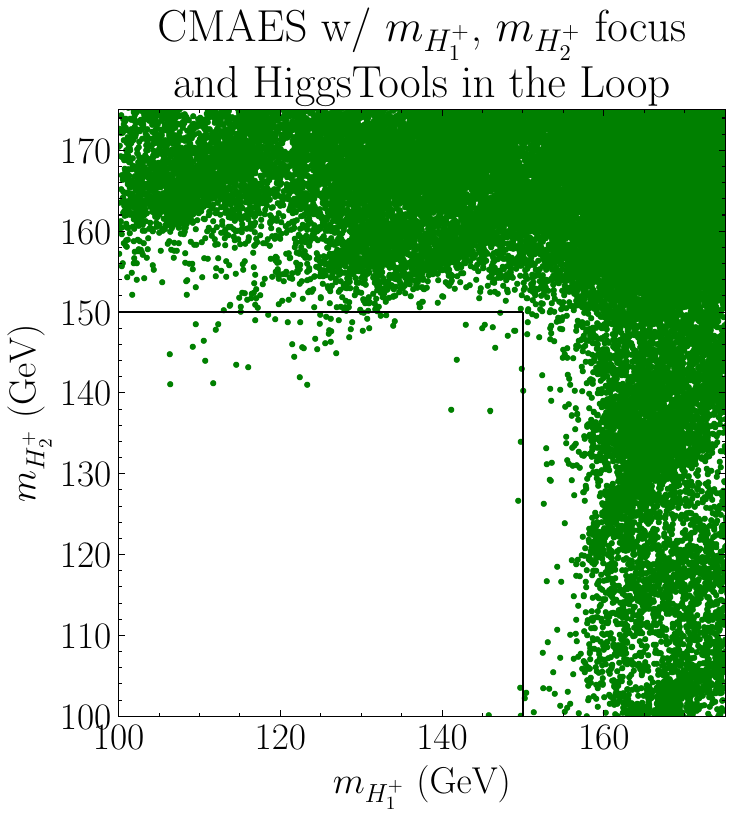}
		\end{subfigure}
	}
	\caption{$(m_{H^+_1}, m_{H^+_2})$ scatter plot zoomed at $m_{H^+_{1,2}} \leq 150$ GeV for different runs. Only points passing \texttt{HiggsBounds} constraints are shown.}
	\label{fig:mC1_mC2_zoomed}
\end{figure}

The above results exhibit a peculiar feature that warrants further discussion. Upon closer inspection of the points that survive \verb|HiggsBounds| when comparing~\cref{fig:mC1_mC2_no_HT} to~\cref{fig:mC1_mC2_w_HT}, we see that the scans without \verb|HiggsTools| in the loop appear to have two \emph{islands} of points at $m_{H^+_{1/2}}\sim 140$ GeV, which the scans with \verb|HiggsTools| in the loop missed. We present a \emph{zoomed} look of this region in~\cref{fig:mC1_mC2_zoomed}, where we only show the points that have passed \verb|HiggsBounds| constraints. This suggests that we have not completely mitigated the excessive \emph{eagerness} of CMAES, which might lead us to miss multimodal solutions, i.e., disjoint valid regions of the parameter space.

\begin{figure}[H]
	\makebox[\textwidth][c]{
		\begin{subfigure}{0.375\textwidth}
			\includegraphics[width=\linewidth]{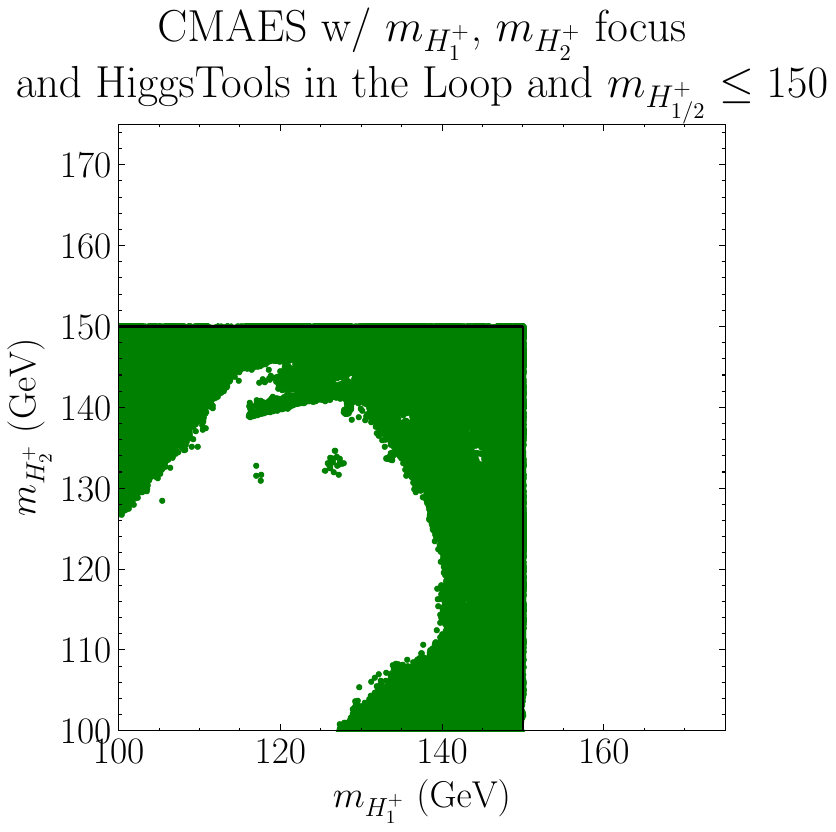}
		\end{subfigure}
	}
	\caption{$(m_{H^+_1}, m_{H^+_2})$ scatter plot for a CMAES run with parameter space restricted to $m_{H^+_{1,2}} \leq 150$ GeV.}
	\label{fig:mC1_mC2_zoomed_scan}
\end{figure}

To better understand whether CMAES is being driven away from this region by its \emph{eagerness} to converge, we performed a dedicated scan where we restricted the parameter space to $m_{H^+_{1,2}} \leq 150$ GeV, and with all other parameter bounds unchanged. We present the result in~\cref{fig:mC1_mC2_zoomed_scan}, where we notice that if restricted to that region, CMAES will explore it extensively. Furthermore, we notice that the points $m_{H^+_{1/2}}\sim 140$ GeV, which seem above to populate two disjoint regions, do not describe isolated~\emph{islands} of the valid parameter space. There are two important conclusions to draw from this. The first conclusion is that the empty regions of the scatter plot of valid points produced by CMAES do not equate to regions without valid points. This means that one has to be very careful when interpreting these~\emph{seemingly empty} regions without studying them in detail. The second conclusion is that when one focusses on studying these regions, one can find a completely different picture than assumed. In this case the previous results, both from alignment limit scan and from CMAES without \verb|HiggsTools| in the loop, suggested that there are multiple disjoint regions of valid points in the $(m_{H^+_1}, m_{H^+_2})$ plane for $m_{H^+_{1/2}}\sim 140$ GeV, whereas a closer inspection teaches us that this is not the case.

\subsection{Rewarding Exploration in the Observable Space\label{subsec:observable_space_reward}}

So far we have shown how we can improve the parameter space coverage by providing CMAES with a novelty reward in the parameter space by turning on a density penalty in observable values $p(\{\OB_i\})$. However, a more interesting avenue is to apply the novelty reward to the observable space, $\OB$,\footnote{We abuse terminology by calling observable all the physical quantities which are constrained. This is not strictly correct, as many constraints are theoretical, and some parameters (namely the masses) are themselves physical observables. The purpose of this section is to study how we can achieve exploration through the constrained quantities and observables.} as this will allow us to assess whether there is new phenomenology obscured by traditional random sampling techniques.

\begin{figure}[htb]
	\makebox[\textwidth][c]{
		\begin{subfigure}[t]{0.375\textwidth}
			\includegraphics[width=\linewidth]{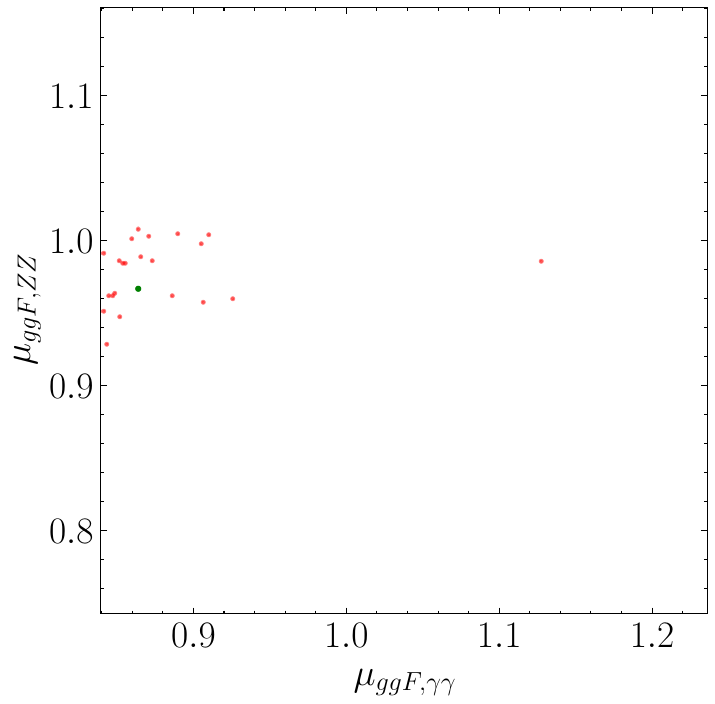}
			\caption{Random sampling}
			\label{fig:mugg_vs_muZZ_no_HT_k9}
		\end{subfigure}
		\begin{subfigure}[t]{0.375\textwidth}
			\includegraphics[width=\linewidth]{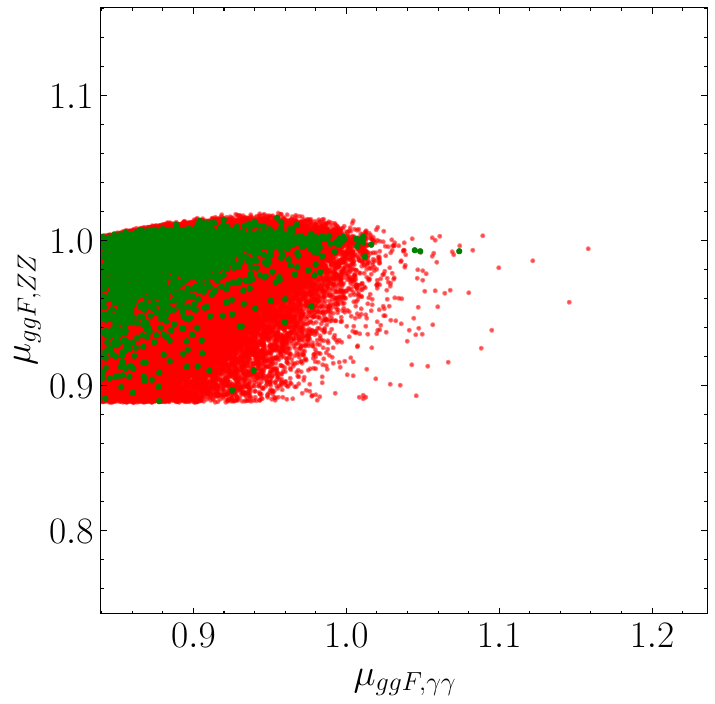}
			\caption{Alignment limit: AL-2}
			\label{fig:mugg_vs_muZZ_no_HT_k16}
		\end{subfigure}
		\begin{subfigure}[t]{0.375\textwidth}
			\includegraphics[width=\linewidth]{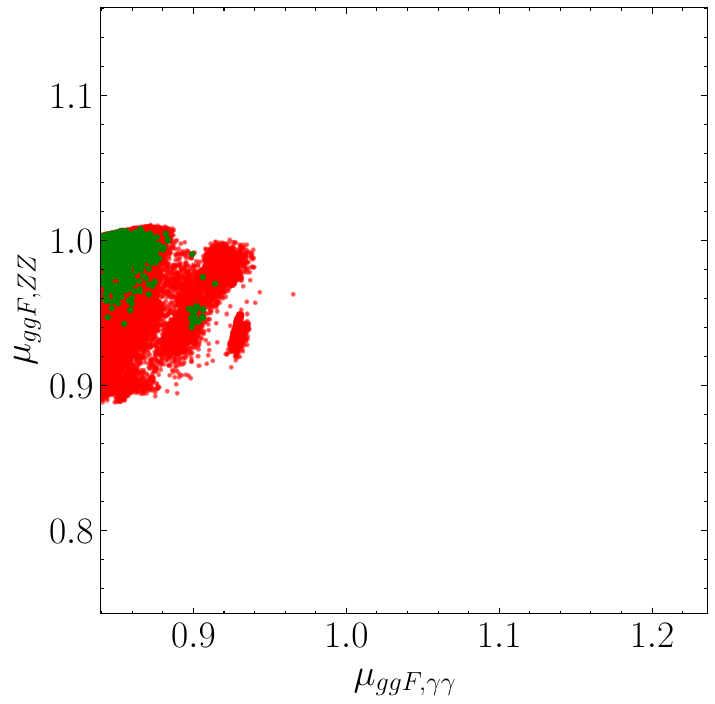}
			\caption{CMAES no penalty}
			\label{fig:mugg_vs_muZZ_no_HT_cmaes}
		\end{subfigure}
	}\\
	\makebox[\textwidth][c]{
		\begin{subfigure}[t]{0.375\textwidth}
			\includegraphics[width=\linewidth]{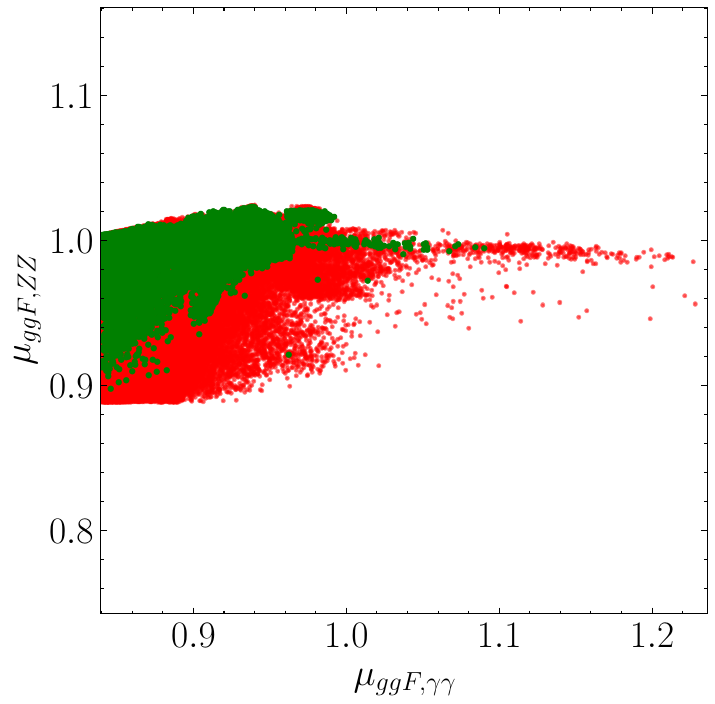}
			\caption{CMAES with parameter novelty reward}
			\label{fig:mugg_vs_muZZ_no_HT_cmaes_parameter_penalty}
		\end{subfigure}
		\begin{subfigure}[t]{0.375\textwidth}
			\includegraphics[width=\linewidth]{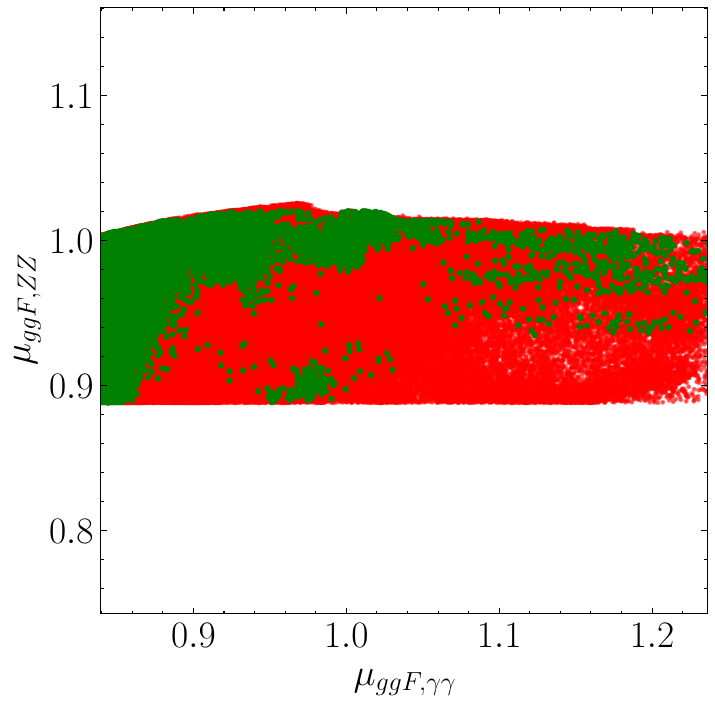}
			\caption{CMAES with observable novelty reward}
			\label{fig:mugg_vs_muZZ_no_HT_cmaes_observable_penalty}
		\end{subfigure}
	}
	\caption{$(\mu_{ggF,\gamma\gamma}, \mu_{ggF,ZZ})$  scatter plot for different runs without novelty reward. A point surviving \texttt{HiggsTools} is represented in green, otherwise in red.}
	\label{fig:mugg_vs_muZZ_no_HT}
\end{figure}

In our first study we want to assess the impact on using a novelty reward in the observable versus the novelty reward in the parameter space studied above. In~\cref{fig:mugg_vs_muZZ_no_HT} we show the $(\mu_{ggF,\gamma\gamma}, \mu_{ggF,ZZ})$ scatter plots for different scans without \verb|HiggsTools| in the loop. Similarly to the previous discussions on parameter space coverage, we see that CMAES without further exploration,~\cref{fig:mugg_vs_muZZ_no_HT_cmaes}, provides a narrower coverage of the observable space than the alignment limit sampling strategy, adding to the intuition that CMAES alone is too eager to converge to be a reliable tool to draw a complete picture of the Physics. This changes considerably once we turn on the parameter space novelty award already studied, which also leads to a greater exploration of the observable space, as can be seen in~\cref{fig:mugg_vs_muZZ_no_HT_cmaes_parameter_penalty}. This is easy to interpret, as forcing CMAES to explore the parameter space will always impact the value of the physical quantities of the model. We notice, however, that this is a byproduct of the parameter space exploration, as in this case CMAES does not have an explicit \emph{incentive} to produce new observable values. In~\cref{fig:mugg_vs_muZZ_no_HT_cmaes_observable_penalty} we show the result of turning on the density penalty in the observable space, therefore explicitly forcing CMAES to find \emph{points with different phenomenology}. The result is stunningly different from all the other scans, with CMAES finding points with a far more diverse value than any of the other scans considered so far. Of particular interest, we observe how CMAES can find points with $\mu_{ggF, \gamma\gamma}\simeq 1.2$, which was completely obscured using alignment limit random sampling and painting a very different picture of what phenomenology the $Z_3$ 3HDM model can have.

\begin{figure}[htb]
	\makebox[\textwidth][c]{
		\begin{subfigure}[t]{0.375\textwidth}
			\includegraphics[width=\linewidth]{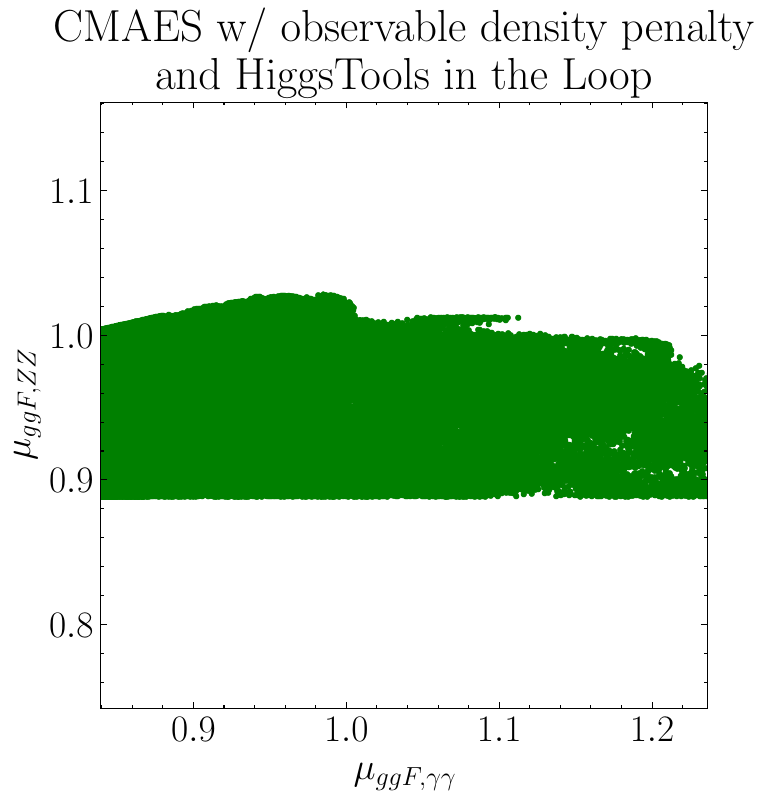}
		\end{subfigure}
	}
	\caption{$(\mu_{ggF,\gamma\gamma}, \mu_{ggF,ZZ})$ scatter plot for CMAES with novelty reward in observable space and \texttt{HiggsTools} constraints included in the loss function.}
	\label{fig:mugg_vs_muZZ_w_HT}
\end{figure}

Having shown how an observable space penalty can drive CMAES exploration into novel phenomenological realisations of the model, we now perform the scan with \verb|HiggsTools| in the loop to endow CMAES with information of the \verb|HiggsBounds| constraints (themselves added to the loss function and to the penalty). The resulting $(\mu_{ggF,\gamma\gamma}, \mu_{ggF,ZZ})$ scatter plot is shown in~\cref{fig:mugg_vs_muZZ_w_HT}, where we observe how CMAES was able to find points across all allowed values (up to 2-$\sigma$ with experimental measurement) for $\mu_{ggF,\gamma\gamma}$ with $0.89 \lesssim \mu_{ggF,ZZ}\lesssim 1.025$, while completely \emph{rediscovering} the possible values produced by the alignment limit sampling strategy. This result highlights the power and versatility of our methodology to find new phenomenological realisations of a model.

\begin{figure}[H]
	\makebox[\textwidth][c]{
		\begin{subfigure}[t]{0.375\textwidth}
			\includegraphics[width=\linewidth]{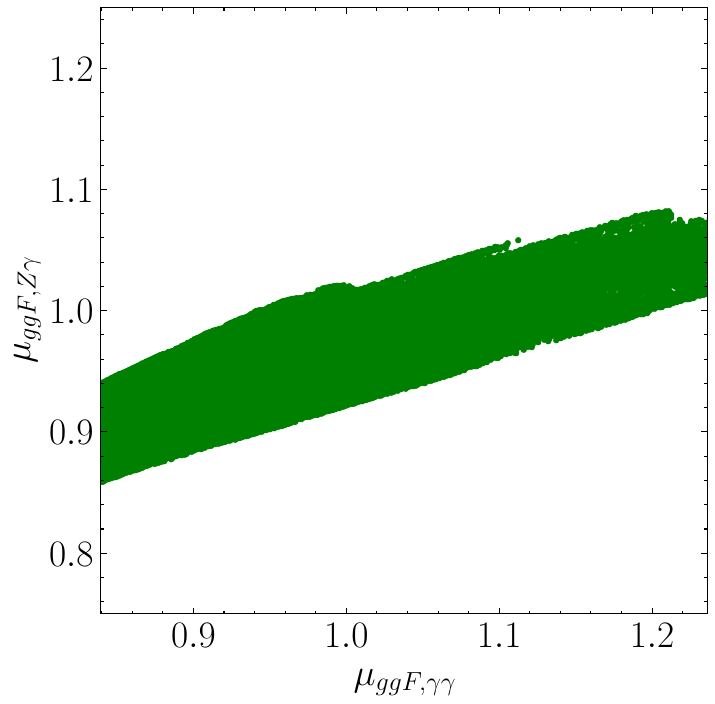}
			\caption{CMAES with observable novelty reward}
			\label{fig:mugg_vs_muZZ_w_HT_observable_penalty}
		\end{subfigure}
		\begin{subfigure}[t]{0.375\textwidth}
			\includegraphics[width=\linewidth]{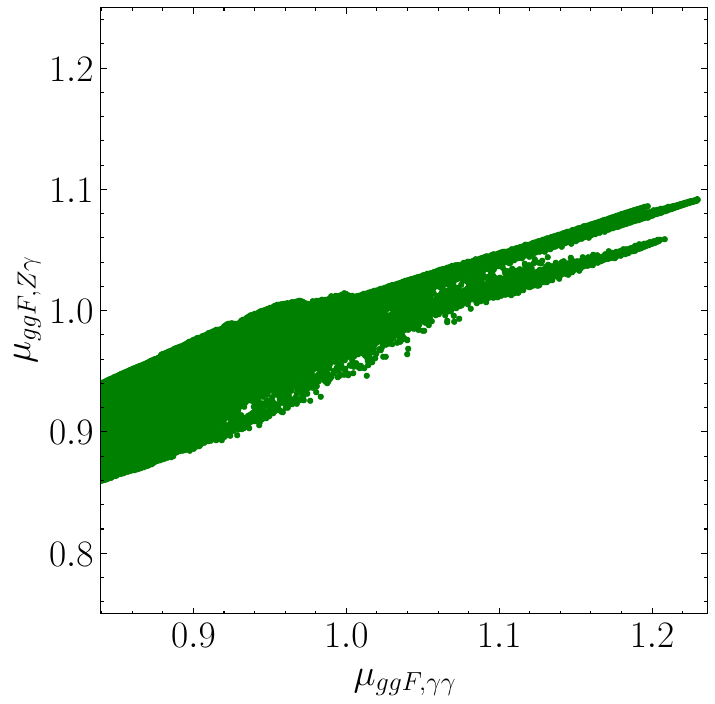}
			\caption{CMAES with $\mu_{ggF,\gamma\gamma}, \mu_{ggF,ZZ}$ focus}
			\label{fig:mugg_vs_muZZ_w_HT_focused}
		\end{subfigure}
	}
	\caption{$(\mu_{ggF,\gamma\gamma}, \mu_{ggF,Z\gamma})$ scatter plot for CMAES with focused novelty reward in observable space and \texttt{HiggsTools} constraints included in the loss function.}
	\label{fig:mugg_vs_mugZ}
\end{figure}

Just like we could \emph{focus} the parameter density on a subset of parameters, we can also focus the observable density penalty on a subset of constraints, allowing one to explore to what extent the model explains certain experimental results. For example, recently~\cite{ATLAS:2023ssf} ATLAS and CMS have released their most recent measurements of the Higgs decaying to $Z\gamma$ with $\mu_{Z\gamma} = 2.2 \pm 0.7$, which is just compatible with the Standard Model within 2-$\sigma$. Then one can ask whether the $Z_3$ 3HDM model discussed in this work could explain such a high value of $\mu_{Z\gamma}$, considering that, without additional states\footnote{To be able to have such a situation on has to go beyond NHDM, see for instance the discussion in Ref.\cite{Boto:2023bpg}.}, the Higgs decay channels are considerably correlated, preventing any particular $\mu_{ij}$ to be large while all others remain small. To study this, we performed a scan with a focused observable density penalty over $(\mu_{ggF,\gamma\gamma}, \mu_{ggF,Z\gamma})$, which we present in~\cref{fig:mugg_vs_mugZ} alongside the scatter plot obtained by the CMAES run with observable density computed over all constraints. Perhaps surprisingly, we see that the scan with the focused density penalty,~\cref{fig:mugg_vs_muZZ_w_HT_focused}, has covered a smaller region than the one with the density penalty computed using all constraints,~\cref{fig:mugg_vs_muZZ_w_HT_focused}. A possible interpretation for this is that the focused density is too constraining, preventing CMAES from finding other ways to populate this plane around other constraints. Conversely, the run with penalty over all constraints will be less demanding for CMAES to explore this subspace, as CMAES will find a way of reducing the penalty by spreading the possible constraint values elsewhere, eventually finding a new route to new values in the $(\mu_{ggF,\gamma\gamma}, \mu_{ggF,Z\gamma})$ plane. In other words, although when projected onto the $(\mu_{ggF,\gamma\gamma}, \mu_{ggF,Z\gamma})$ plane the valid region appears simply connected, the overall geometry and topology of the valid region of the parameter space are likely far more intricate with focused scans obfuscating these nuances. This interplay between a focused density, the availability of paths for CMAES to explore, and the topological and geometrical details of the valid region is an aspect of our methodology that will be further explored in future work.

\subsection{Using Points as Seeds for New Runs\label{subsec:seed_points}}

The scans performed so far have highlighted the versatility of our methodology in exploring parameter (and observable) spaces. However, the runs performed are independent of each other, i.e., while each run has its own parameter/observable density estimator, this is only trained using valid points found during that run alone.

Then, one can entertain the idea of reusing the information of previous scans to guide new runs in regions of interest. In this section, we explore this idea and provide an example of its implementation by choosing valid points from the previous scans as a \emph{seed} to new runs. Recalling that CMAES can be initiated with an explicit mean, i.e. starting position, and an overall scale of the covariant matrix, $\sigma$, we can then use a valid point as the starting position of a new scan. In order to start exploring the vicinity of our starting position $\sigma$ cannot be too large, and we found that setting it to $\sigma=0.01$ guarantees that CMAES starts already at the minimum of the constraint loss function.

\begin{figure}[H]
	\makebox[\textwidth][c]{\includegraphics[width=1.2\linewidth]{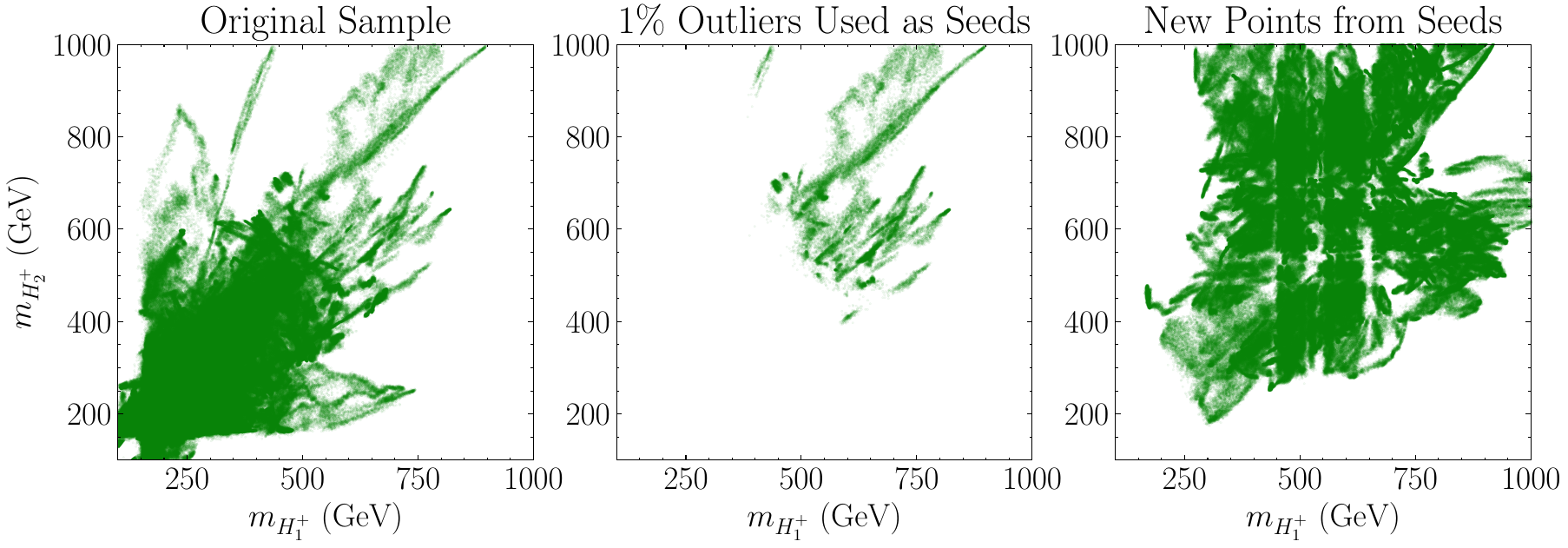}}
	\caption{$(m_{H^+_1}, m_{H^+_2})$  scatter plot for the seeded run. Left: The whole collection of valid points obtained by the other scans. Middle: The $1\%$ outliers classified by HBOS. Right: New points obtained by the seeded runs started at points randomly selected from the $1\%$ outliers.}
	\label{fig:mC1_mC2_seeded_runs}
\end{figure}

Seed points were identified by running HBOS on the entire collection of valid points (left pane of~\cref{fig:mC1_mC2_seeded_runs}). For this concrete example, we evaluated the density only on the $(m_{H^+_1}, m_{H^+_2})$ subspace and identified the $1\%$ outliers (middle pane of~\cref{fig:mC1_mC2_seeded_runs}), i.e., the points representing the least explored parts of the valid region of the parameter space. We notice some of the shortcomings of HBOS as the density estimator in this plot: given that HBOS fits a histogram to each dimension to compute the density, a point might be in a relatively sparse region but might not be picked as an outlier if one of its components is in a populated bin. For example, we see that the outliers are not necessarily at the rim (convex haul) of the space, but in regions where there are few points both in $m_{H^+_1}$ and $m_{H^+_2}$. This same shortcoming of HBOS is present in the scans with novelty reward, leaving room for improvement to be explored in future work.

With the most outlying valid points identified, we ran $100$ scans not only seeded by a point randomly chosen from the $1\%$ outlier subset, but with the density penalty also making use of the outliers to guide the new scans away from the already explored regions. We did not use the whole sample of valid points to train the density estimator as it is comprised of over 4 million valid points, considerably slowing down the scan. In the right pane of~\cref{fig:mC1_mC2_seeded_runs} we show the resulting valid points found by the seeded scans, where we observe that CMAES was able to explore even further away from the previously chartered valid region. Clearly, one could now use the new points as new seeds in repeated iterations to explore even more this subsection of the parameter space, or any other section of it or of the observable space, in order to draw an even more global picture of the valid region. The caveat of only using \emph{chained} scans is that one can only explore regions that are simply connected to the seed, a detail which must be kept in mind when employing this strategy. Finally, we notice that the scatter plot appears to have some vertical and horizontal regions with less points, this is an artefact of HBOS which draws a histogram with $100\times100$ bins in this subspace, impacting the density value along horizontal and vertical strips with width similar to the width of the bins.

\subsection{Convergence Metrics\label{sec:convergence_metrics}}

\begin{figure}[H]
	\makebox[\textwidth][c]{\includegraphics[width=0.6\linewidth]{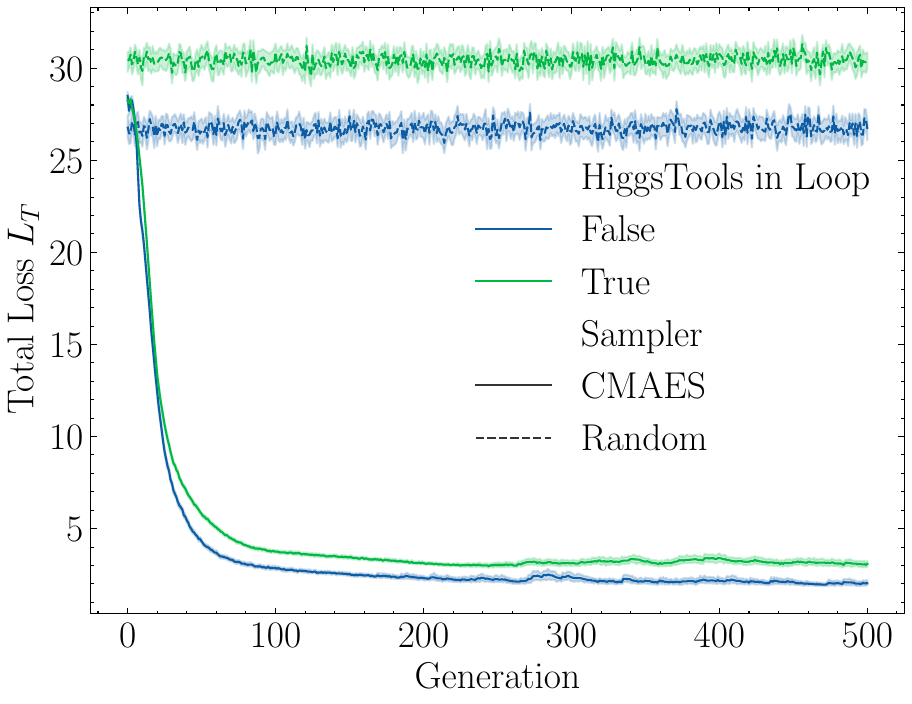}}
	\caption{Total loss as function of generation. Only the first 500 generations are shown.
			{The solid (broken) lines are for the sampler CMAES (Random) while the color green (blue) indicates whether HiggsTools was used (not used) in the loop.}
		The random sampler curves are over random generations of $12$ poitns, the same population size as CMAES. The shaded regions represent $0.95$ confidence intervals computed using a bootstrap of 100 runs.}
	\label{fig:mean_constraints_HT_vs_noHT}
\end{figure}

Having discussed how density penalties can be used to enhance the CMAES exploration of the parameter space, we now turn to another aspect of our methodology: the convergence speed. Recalling that CMAES operates by minimising the total loss function, $L_T$ from~\cref{eq:loss}, we show how its value decreases sharply after just a few generations in~\cref{fig:mean_constraints_HT_vs_noHT}, where we also provide the values of random generations for comparison. More precisely, after just 100 generations, totalling around just 1200 points, CMAES has nearly converged to the valid region of the parameter space.

Despite the suggestive previous plot, not all scans converge within the budget, which we set to 100 runs of 2000 generations for each case in~\cref{tab:scans} without \verb|HiggsTools|, and to 200 runs of 2000 generations for the cases with \verb|HiggsTools|.\footnote{One could alternatively increase the budget for the \texttt{HiggsTools} by increasing the number of generations to 4000. Intuitively, more runs provide a more global picture, whereas longer runs allow for longer explorations of the valid region. The choice between allocating more budget to one over the other depends on the intended study.} We present these metrics in~\cref{tab:scan-statistics}, {where in the last two columns we show} the fraction of the {obtained} valid points that are within {each of the} alignment cases, AL-1 from~\cref{Al-1} and AL-2 from~\cref{Al-2}. We see that, while most points are within AL-1, only a minority are within AL-2. More interestingly, we observe how the scan with novelty reward in the $\alpha_1,\ \beta_1,\ \alpha_2,\ \beta_2$ subspace of the parameter space has produced the most points away from the alignment limit. This can be visually understood in~\cref{fig:sa1b1_vs_sa2b2_no_HT} and~\cref{fig:sa1b1_vs_sa2b2_w_HT} where it is clear that the novelty reward is guiding CMAES to values of $\alpha_1,\ \beta_1$ that are beyond the alignment limit bounds. This is a feature of the versatility of our methodology, as we can perform dedicated scans explicitly away from previously considered priors and regions of the parameter space.
\begin{table}[H]
	\makebox[\textwidth][c]{
		\begin{tabular}{p{2.9cm}p{4.6cm}p{3.1cm}p{2.3cm}p{2.3cm}}
			\hline\hline
			Sampling                                & Scan                                                  & Converged Runs & Within AL-1 & Within AL-2        \\ \hline
			\multirow{6}{*}{CMAES}                  & No penalty                                            & 97 out of 100  & $1.00$      & $1.5\times10^{-2}$ \\
			                                        & Parameter novelty reward                              & 95 out of 100  & $0.90$      & $5.1\times10^{-3}$ \\
			                                        & Observable novelty reward                             & 90 out of 100  & $0.99$      & $1.8\times10^{-4}$ \\
			                                        & $\alpha_1$, $\beta_1$, $\alpha_2$, $\beta_2$ focus    & 94 out of 100  & $0.85$      & $5.9\times10^{-3}$ \\
			                                        & $\mu_{ggF, \gamma\gamma}$, $\mu_{ggF, Z\gamma}$ focus & 91 out of 100  & $1.00$      & $4.2\times10^{-5}$ \\
			                                        & $m_{H_1^+}$, $m_{H_2^+}$ focus                        & 90 out of 100  & $0.95$      & $6.6\times10^{-3}$ \\ \hline
			\multirow{5}{*}{CMAES}                  & Parameter novelty reward                              & 92 out of 200  & $0.92$      & $1.1\times10^{-2}$ \\
			\multirow{5}{*}{w/ \texttt{HiggsTools}} & Observable novelty reward                             & 111 out of 200 & $0.94$      & $1.4\times10^{-2}$ \\
			                                        & $\alpha_1$, $\beta_1$, $\alpha_2$, $\beta_2$ focus    & 102 out of 200 & $0.84$      & $4.1\times10^{-2}$ \\
			                                        & $\mu_{ggF, \gamma\gamma}$, $\mu_{ggF, Z\gamma}$ focus & 101 out of 200 & $0.93$      & $4.1\times10^{-3}$ \\
			                                        & $m_{H_1^+}$, $m_{H_2^+}$ focus                        & 91 out of 200  & $0.98$      & $5.3\times10^{-2}$ \\ \hline\hline
		\end{tabular}
	}
	\caption{\label{tab:scan-statistics}Convergence and coverage statistics of the different scans presented in~\cref{tab:scans}. {The last two columns represent the fractions of valid points obtained by each of the scans that are within both alignment limits.} While the first CMAES rows correspond to scans without \texttt{HiggsTools} in the loop, the corresponding fractions of points in both alignment cases are computed using points that have passed \texttt{HiggsBounds} constraints after the scan.}
\end{table}

\begin{figure}[H]
	\makebox[\textwidth][c]{\includegraphics[width=0.6\linewidth]{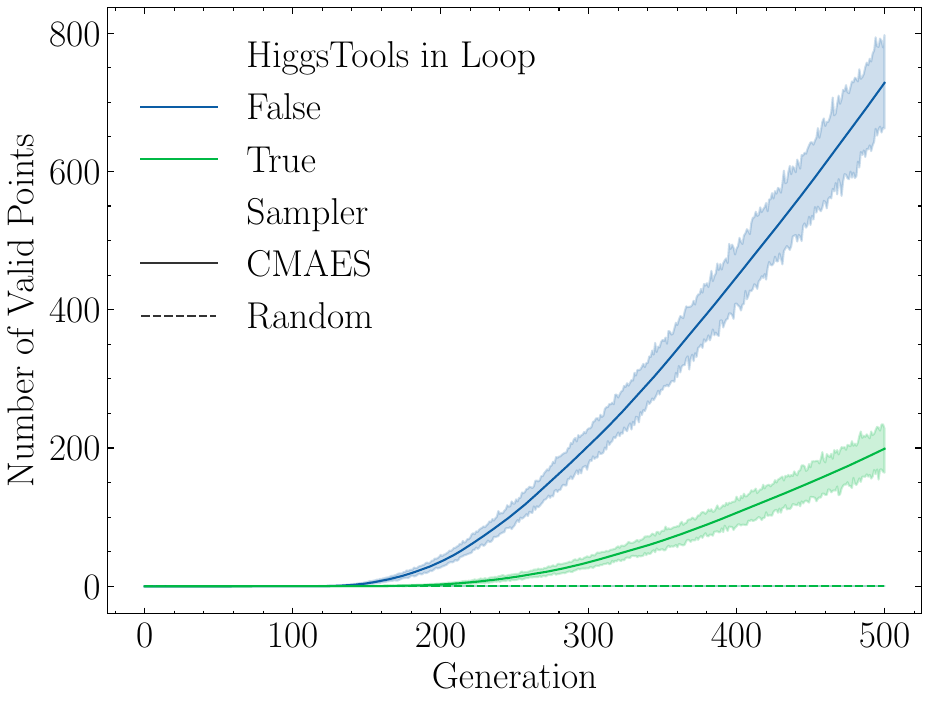}}
	\caption{Number of valid points founds as function of generation. Only the first 500 generations are shown.
			{The solid (broken) lines are for the sampler CMAES (Random) while the color green (blue) indicates whether HiggsTools was used (not used) in the loop.}
		The random sampler curves are over random generations of $12$ points, the same population size as CMAES. The shaded regions represent $0.95$ confidence intervals computed using a bootstrap of 100 runs.}
	\label{fig:n_valid_points_HT_vs_noHT}
\end{figure}

As CMAES converges, it will start to find good points. This can be seen
in~\cref{fig:n_valid_points_HT_vs_noHT} where we observe that after around 100 generations for the runs without \verb|HiggsTools| in the loop, and after around 200 generations for the runs with \verb|HiggsTools| in the loop, CMAES starts finding valid points.

\begin{table}[H]
	\makebox[\textwidth][c]{
		\begin{tabular}{p{2cm}p{3cm}p{2cm}p{3cm}p{3cm}}
			\hline\hline
			       & \multicolumn{2}{c}{Before \texttt{HiggsTools}} & \multicolumn{2}{c}{After \texttt{HiggsTools}}                                                       \\
			       & Points                                         & Efficiency                                    & Points                   & Efficiency               \\ \hline
			AL-1   & 21 in $4.4 \times 10^{12}$                     & $\mathcal{O}(10^{-11})$                       & 0 in $4.4 \times10^{12}$ & $<\mathcal{O}(10^{-12})$ \\
			AL-2   & 13 701 in $10^{10}$                            & $\mathcal{O}(10^{-6})$                        & 510 in $\times10^{10}$   & $\mathcal{O}(10^{-8})$   \\
			Random & 23 in $ 10^{13}$                               & $\mathcal{O}(10^{-12})$                       & 1 in $ 10^{13}$          & $\mathcal{O}(10^{-13})$  \\
			\hline\hline
		\end{tabular}
	}
	\caption{Sampling efficiencies of random sampling strategies. We note that for the completely random scan the numbers are estimated.}
	\label{tab:efficiency}
\end{table}

\begin{figure}[H]
	\makebox[\textwidth][c]{\includegraphics[width=0.6\linewidth]{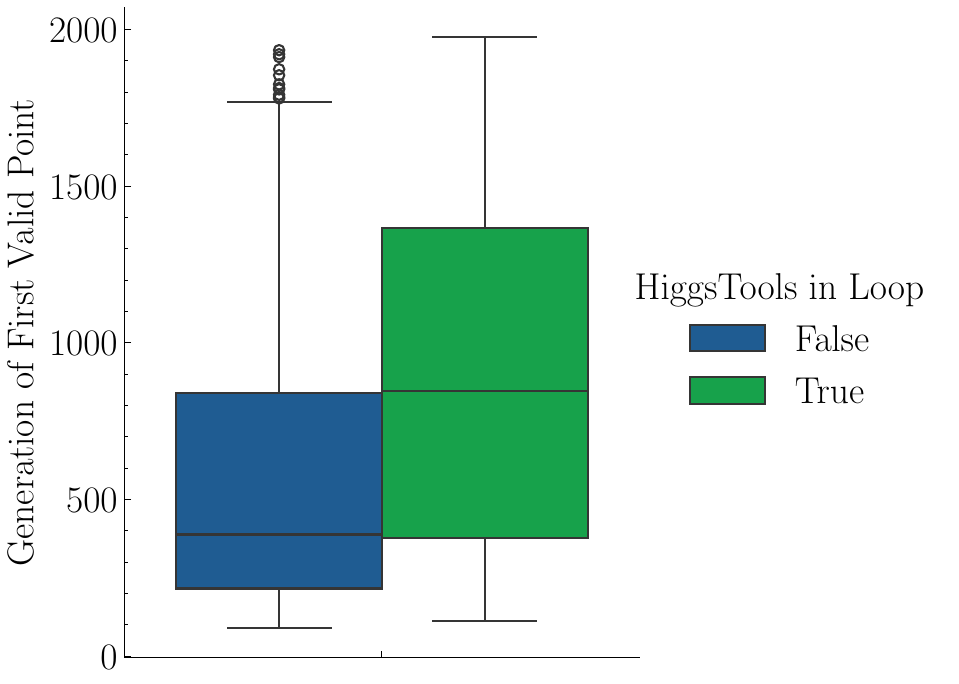}}
	\caption{Distribution of the generation with the first valid point. The distributions shown are obtained from all runs with and without \texttt{HiggsTools} in the loop.}
	\label{fig:generation_first_valid_point_HT_vs_noHT}
\end{figure}

In~\cref{fig:generation_first_valid_point_HT_vs_noHT} we {use a boxplot to represent} the distribution of the {value of the} generation {number when} the first valid point was {obtained} for the runs with and without \verb|HiggsTools| in the loop. We see that checking for \verb|HiggsBounds| \emph{postpones} the discovery of the first good point by a factor of around $2$ in terms of generation number. However, this is far better than with random sampling (both purely random and around the alignment limit) where only at most $5\%$ of the points that respect all other constraints survive \verb|HiggsBounds| constraints,~\cref{tab:efficiency}. Additionally, we note that the number of trial points needed to find a good valid point is around $\mathcal{O}(250-750)\times 12$ when not including \verb|HiggsBounds| constraints in the loop, and $\mathcal{O}(400-1300)\times 12$ when including \verb|HiggsBounds| constraints in the loop. Recall that the efficient sampling of the random sampler is (estimated) to be $\mathcal{O}(10^{-12})$ and $\mathcal{O}(10^{-13})$ respectively, which means that our methodology improves the sampling efficiency by a factor of 8 orders of magnitude, even if we do not consider the near perfect sampling efficiency after convergence during the density penalty-guided exploration phase. Unsurprisingly, the improvement over the AL-2 alignment limit sampling strategy is more modest, with around four orders of magnitude when considering \verb|HiggsBounds| constraints, but we reiterate that this only pertains to the first convergence and that our algorithm can then continue to explore the region found and go beyond the alignment limit bounds, as was highlighted in the previous section.

\begin{figure}[H]
	\makebox[\textwidth][c]{\includegraphics[width=0.6\linewidth]{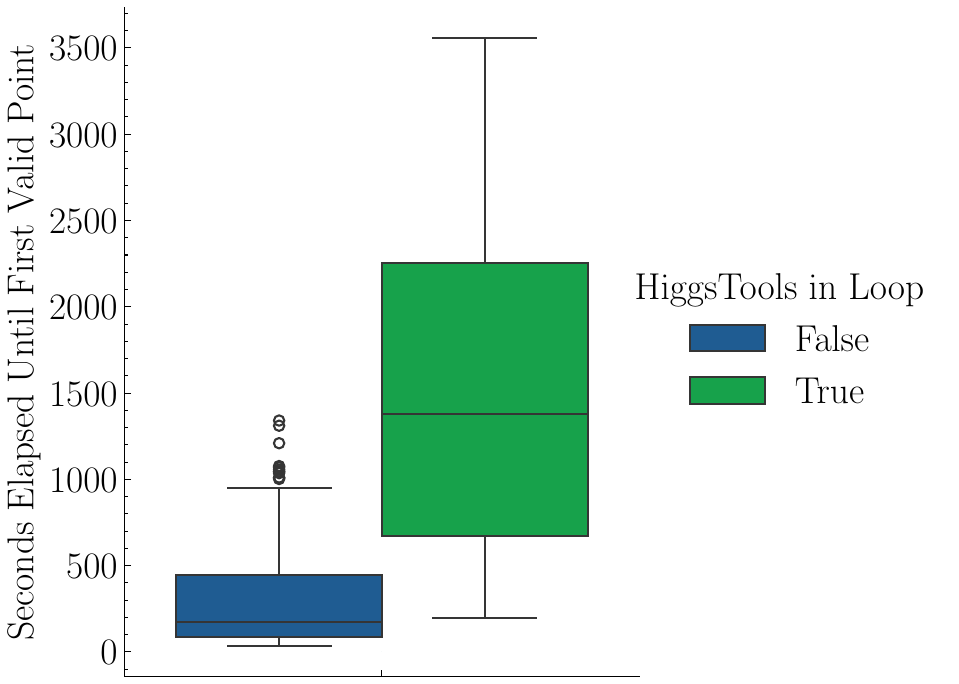}}
	\caption{Distribution of the elapsed time until the first valid point. The distributions shown are obtained from all runs with and without \texttt{HiggsTools} in the loop.}
	\label{fig:time_elapsed_first_valid_point_HT_vs_noHT}
\end{figure}

So far, we have seen the improvements to convergence provided by CMAES by taking into account the number of points tried, observing massive improvements over random sampling. On the other hand, the methodology presented in this work is only useful if it also provides a speed-up in terms of wall time, i.e. time passed from the reference frame of the user. In~\cref{fig:time_elapsed_first_valid_point_HT_vs_noHT} we present a variation of~\cref{fig:generation_first_valid_point_HT_vs_noHT} but in terms of elapsed time, instead of generations. We see that CMAES without including \verb|HiggsBounds| constraints in the loop tends to find points within $\mathcal{O}(100-500)$ seconds, i.e. in minutes, while when including \verb|HiggsBounds| constraints, this increases do $\mathcal{O}(750-2250)$ seconds. The slowing down is easily understood: using \verb|HiggsTools| slows down the evaluation of a point by a factor of around 3 (see below for more details), and since converging on \verb|HiggsBounds| constraints delays CMAES in finding a good point by a factor of around 2 (see discussion above) we expect an overall wall time delay of 5, which is what we can see here.

To better understand the impact of the different components of our methodology in the total time, we present in~\cref{tab:times} the times taken by different steps of the loop for CMAES and the random sampler, and with and without \verb|HiggsTools| in the loop. In this table, generation time represents the time needed to perform all the steps of a generation including evaluation time, i.e. the time needed to compute all observables (including \verb|HiggsBounds|, when applicable), train the density estimator (when applicable), and perform diverse housekeeping tasks such as save intermediate results, keep track of run metrics, etc. In this table we see that the overall housekeeping overhead can be assessed in the random sampler rows as for these there is no overhead related to CMAES and to density estimation, and it is around 0.015 (0.019) seconds without (with) \verb|HiggsTools|.\footnote{The larger housekeeping overhead associated with \texttt{HiggsTools} is due to the presence of more metrics to keep track and larger intermediate files to save.} The most important observation to take from this table is that the overall overheard of our methodology, including that associated with CMAES and density estimation, is at most around $10\%$ of the total generation time for the CMAES runs without \verb|HiggsTools|. Once we include \verb|HiggsTools| in the loop, the overall generation time increases 3-fold but the overhead remains mostly the same, showing that our methodology provides even greater gains for problems with a slow evaluation time. Additionally, we see that our choice of HBOS for density estimator corresponds to a minor fraction of the overhead. However, one can notice that the standard deviation of the density estimator training is greater than the mean; this is because HBOS has a linear computational complexity with respect to the number of valid points, effectively becoming slower to train the more valid points we have found. Improvements to the density estimator are left for future work.
\begin{table}[H]
	\makebox[\textwidth][c]{
		\begin{tabular}{p{2cm}p{2.2cm}p{2.5cm}p{2.5cm}p{3.25cm}p{2.8cm}}
			\hline\hline
			Sampler                 & \texttt{HiggsTools} in Loop & Generation      & Evaluation      & Density Estimator            & Overhead          \\\hline
			\multirow{2}{*}{CMAES}  & False                       & $0.48 \pm 0.16$ & $0.43 \pm 0.15$ & $(3.9 \pm 23)\times 10^{-3}$ & $0.055 \pm 0.043$ \\
			                        & True                        & $1.6 \pm 0.3$   & $1.6 \pm 0.2$   & $(1.9 \pm 22)\times 10^{-3}$ & $0.068 \pm 0.048$ \\
			\multirow{2}{*}{Random} & False                       & $0.33 \pm 0.06$ & $0.31 \pm 0.06$ & N/A                          & $0.015 \pm 0.003$ \\
			                        & True                        & $1.4 \pm 0.2$   & $1.3 \pm 0.2$   & N/A                          & $0.019 \pm 0.012$ \\ \hline\hline
		\end{tabular}
	}
	\caption{Comparison of times, in seconds, taken by different parts of the loop for CMAES and random sampler with and without \texttt{HiggsTools} in the loop. The times refer to a generation of 12 points in every case. The values presented are the mean $\pm$ one standard deviation over the scans falling in each of the four categories.}
	\label{tab:times}
\end{table}


\section{Conclusions}
\label{s:conclusions}

In this paper, we have developed a novel approach to explore the highly constrained multidimensional parameter space of the $Z_3$ 3HDM, defined in~\cref{s:model} and constrained discussed in~\cref{s:constraints} and~\cref{s:appendix_A}, and go beyond alignment limit priors presented in~\cref{s:scan}, by combining CMAES power of exploration with a Machine Learning estimator for point density.

It is important to note that, while the subject of study in this paper was the $Z_3$ 3HDM parameter space, our approach is general and applicable to any Physics case, providing a solution to the difficulty of sampling good points in highly constrained multidimensional parameter spaces.

In~\cref{s:AI} we introduced our strategy, using CMAES, a powerful evolutionary strategy, in combination with HBOS, a fast ML model for density estimation. Our approach guarantees that the density-based \emph{novelty reward} does not compete in the loss function with the constraints on the model and pushes CMAES to explore the parameter space once converged. Importantly, we showed how our methodology is versatile, as we can turn on the \emph{novelty reward} in the parameter space or in the observable space, where the phenomenology is realised. Additionally, the \emph{novelty reward} can be computed by estimating the density in only a subset of parameters and/or observables, allowing for quick focused scans on regions of interest.

In~\cref{s:results} we presented the results of multiple scans performed with our methodology, each with different combinations of parameters and/or observables on which the density penalty was computed. We showed how our approach can effortlessly go beyond the alignment limit sampling strategies, finding valid points in regions of the parameter space hitherto ignored by such sampling strategies. More precisely, using the novelty reward in the parameter space, both in whole or in a subset, in~\cref{subsec:parameter_space_reward} we have found that it is easy to go beyond the alignment limit in the $(\alpha_1, \beta_1)$ plane. In the same analysis, we showed how our methodology also exposes regions of heavy scalar masses, even preferring it over the region excluded by \verb|HiggsBounds|, which we explored in detail in~\cref{subsubsec:mC1_mC2} by restricting the parameter space to $m_{H^+_{1,2}} \leq 150$ GeV, finding a considerably different picture of that region of the parameter space that one would get from alignment limit sampling.

While we set ourselves to explore the parameter space with CMAES
combined with a~\emph{novelty reward}, the Physics of the model
resides its space of observables and physical
quantities. In~\cref{subsec:observable_space_reward} we have the results of
scans where the density penalty was computed in the observable space
instead of the parameter space. The results uncover novel possible
phenomenological realisations of the 3HDM, an important contribution
of this work that would not have been possible to achieve without our
AI-based scan. In particular, we find that it is possible to
accommodate Higgs decay signal strengths larger than one up to their
current upper experimental bounds, a phenomenological signature not
captured by alignment limit random sampling strategies. Given the
versatility in exploring different observables, we set to study
whether the $Z_3$ 3HDM can explain the recent measurement of
$\mu_{Z\gamma}\simeq 2.2$ by ATLAS and CMS~\cite{ATLAS:2023ssf},
finding that $\mu_{Z\gamma}\lesssim 1.1$ in the $Z_3$ 3HDM, not a
surprising result given that decay signal strengths are highly
correlated in this model and one cannot get arbitrarily high values
for one of them without spoiling the experimental measurements of the
remainder (for other possibilities, see, for instance, the discussion
in \cite{Boto:2023bpg}). Finally, in~\cref{sec:convergence_metrics} we discuss the convergence metrics of our algorithm and compare it with the pure and alignment limit random sample strategies. Our methodology is orders of magnitude faster (both in number of tried points and wall time) than the random sampling strategies, providing a solution to the random sampling efficiency problem in highly constrained multidimensional parameter spaces.

Although the methodology presented in this work provides impressive speedup and efficiency improvement when compared to random sampling strategies, we have encountered some shortcomings and less appealing characteristics that we want to improve in future work. First, the methodology used in the analyses employs independent runs, each with its own density estimator from which the \emph{novelty reward} is derived. An alternative approach is to share information across runs to ensure the novelty of the exploration. In~\cref{subsec:seed_points} we showed how such a strategy could be implemented, where we identified the valid region of the parameter space less populated (in the $(m_{H^+_1}, m_{H^+_2})$ subspace) by our scans and then used some of the points in that region as a seed for new runs. The resulting new points were significantly different from the ones found previously, highlighting the potential for even further exploration by chaining runs together, a methodological detail that can be improved in the future. Second, in the same study, we encountered some artefacts arising from the \emph{binning} nature of the HBOS, which can, in principle, be mitigated by using a different density estimator (or a different novelty detector). In our early exploration, we tried a variety of alternatives, all significantly slower than HBOS, making our methodology impractical. Producing a better way of assigning the \emph{novelty reward} could solve the binning problem and any manifestation of the \emph{curse of dimensionality} produced by it. Third, we have observed that our methodology might not explore all possible regions as CMAES intuitively follows a \emph{path of fastest descent}. This was particularly clear in~\cref{subsubsec:mC1_mC2} where we addressed the overlooked region of small charged scalar masses. By restricting the parameter space, we were able to populate that region easily, but the fact that it was not explored in the first place shows that we need to be careful when interpreting \emph{empty} regions as regions without valid points. Lastly, we have observed that the geometrical and topological details of the valid region of the parameter space might impact the possible exploration paths of CMAES. This can have a profound effect on the results when there are disjoint, not simply connected, regions of the parameter space supporting good points. We leave to future work the development of a way to assess whether the scans are capable of capturing multimodal valid regions confidently.

Finally, our methodology opens up the possibility for a complete exploration of other $N$HDM (or any other BSM Physics) parameter spaces in light of the current highly constraining experimental results and theoretical conditions. As our work shows, this could lead to novel phenomenological realisations of these models and, ultimately, to the possibility of novel experimental signatures. We leave this phenomenological study for the future.

\section*{Acknowledgments}
\noindent
We would like to welcome to this world and dedicate this work to Leo David Nascimento Crispim Rom\~ao.
We are very thankful to Fernando Abreu de Souza, Nuno Filipe Castro, Andreas Karle, and Werner Porod
for valuable and fruitful discussions.
MCR is supported by the STFC under Grant No.~ST/T001011/1. MCR thanks the Southampton HEP group for the hospitality and access to the infrastructure.
This work is supported in part by the Portuguese Funda\c{c}\~{a}o para
a Ci\^{e}ncia e Tecnologia\/ (FCT) under Contracts
CERN/FIS-PAR/0002/2021, UIDB/00777/2020, and UIDP/00777/2020, these
projects are partially funded through POCTI (FEDER), COMPETE, QREN,
and the EU.

\appendix

\section{Description of the various constraints}
\label{s:appendix_A}

In this appendix, we summarise the constraints that have to be
satisfied for a point in parameter space to be considered a valid
point. As these have already been discussed in great detail in a
series of papers~\cite{Boto:2021qgu,Boto:2022uwv,Boto:2023nyi}, here we just
give a brief review and indicate the places where to look for further
information. We list the constraints in the order in which they are applied in
the code.

\subsection{The $\kappa$'s formalism}

We found that it is useful to select points that are already close to
the LHC constraints, using the $\kappa$'s formalism. We require them
to be within 3$\sigma$ of the LHC data \cite{ATLAS-CONF-2018-031}. The
expressions for the $\kappa$'s for the different types of fermion
couplings in the 3HDM are given in \cite{Boto:2022uwv}. As in this
work we just consider Type I, we have for the fermions,
\begin{equation}
	\kappa_U= \frac{\sin(\alpha_2)}{\sin(\beta_2)},
	\quad
	\kappa_D= \frac{\sin(\alpha_2)}{\sin(\beta_2)},
	\quad
	\kappa_L=  \frac{\sin(\alpha_2)}{\sin(\beta_2)}.
\end{equation}
The couplings with the vector bosons give, for all types,
\begin{equation}
	\label{eq:1a}
	\kappa_W=\cos(\alpha_2) \cos(\alpha_1-\beta_1) \cos(\beta_2) +
	\sin(\alpha_2) \sin(\beta_2)\, ,
\end{equation}
which gives $\kappa_W=1$ when $\alpha_1=\beta_1$ and
$\alpha_2=\beta_2$. We should note that the points are subsequently
tested for the signal strengths, so this constraint is applied with a
large interval (3$\sigma$) just to make the selection faster.

\subsection{Bounded From Below (BFB)}

The scalar potential has to be BFB. As explained in
ref.\cite{Boto:2022uwv}, to find the necessary and sufficient
conditions for this to happen is a difficult task. For the 3HDM is
only known for a few cases with high symmetry in the potential. For
the $Z_3$ 3HDM that we consider here, the best we can do is to use
sufficient conditions. We refer to ref.\cite{Boto:2022uwv} for the
details of the implementation.

\subsection{Oblique parameters $S,T,U$}

To discuss the effect of the electroweak precision parameter,
S, T and U, we use the
expressions in \cite{Grimus:2007if}
and the experimental summary in \cite{Baak:2014ora,Workman:2022ynf}. The expression
for the needed matrices $V$ ($3\times 6$) and $U$ ($3\times 3$)
is~\cite{Boto:2022uwv},
\begin{equation}
	V= \begin{pmatrix} i \textbf{P}^T_{11} & \textbf{R}^T_{11} &
                \textbf{R}^T_{12}   & \textbf{R}^T_{13} & i \textbf{P}^T_{12} & i
                \textbf{P}^T_{13}                                                 \\
                i \textbf{P}^T_{21} & \textbf{R}^T_{21} &
                \textbf{R}^T_{22}   & \textbf{R}^T_{23} & i \textbf{P}^T_{22} & i
                \textbf{P}^T_{23}                                                 \\
                i \textbf{P}^T_{31} & \textbf{R}^T_{31} &
                \textbf{R}^T_{32}   & \textbf{R}^T_{33} & i \textbf{P}^T_{32} & i
                \textbf{P}^T_{33}\end{pmatrix} ,
\end{equation}
and
\begin{equation}
	U=\textbf{Q}^T ,
\end{equation}
where the matrices $\textbf{R},\textbf{P},\textbf{Q}$ were defined
before.

\subsection{Unitarity}

A valid point in the parameter space must also satisfy the perturbative
unitarity constraints.
	{
		These can be expressed in terms of constraints on the $\lambda_i$ parameters of the potential. For this end we require that the eigenvalues of all the two-by-two scattering matrices are bounded by partial wave unitarity. In~\cite{Bento:2017eti} an algorithm was proposed to efficiently obtain the scattering matrices.
		For the different
		symmetry constrained 3HDM these are fully given in~\cite{Bento:2022vsb} to which we refer the reader for further details.
	}

\subsection{The signal strengths $\mu_{ij}$}

The LHC results on the 125 GeV Higgs boson are normally given by the
signal strengths,
\begin{eqnarray}
	\mu_{if} =\left(\frac{\sigma_i^{\text{3HDM}}(pp\to h)
	}{\sigma_i^{\text{SM}}(pp\to
		h)}\right)\left(\frac{\text{BR}^{\text{3HDM}}(h\to
	f)}{\text{BR}^{\text{SM}}(h\to f)}\right) \,,
	\label{e:ss}
\end{eqnarray}
where the subscript `$i$' denotes the production mode and the subscript `$f$'
denotes the decay channel of the SM-like Higgs scalar. The relevant production
mechanisms include gluon fusion~($ggF$), vector boson fusion~($VBF$),
associated production with a vector boson ($VH$, $V = W$ or $Z$), and
associated production with a pair of top quarks ($ttH$). The SM cross
section for the gluon fusion process is calculated using HIGLU
\cite{Spira:1995mt}, and for the other production mechanisms we use
the prescription of Ref.~\cite{deFlorian:2016spz}. The calculated
$\mu_{if}$ are required to be within 2$\sigma$ of the LHC
results\cite{ATLAS:2022vkf}.

\subsection{Constraints from flavour data}

In Type-I, by construction,
3HDM there are no FCNCs at the tree-level. Therefore, the only NP contribution at
one-loop order to observables such as $b\to s\gamma$ and the neutral meson mass
differences will come from the charged scalar Yukawa couplings. We follow~\cite{Chakraborti:2021bpy} where it was shown that
the constraints coming from the meson mass
differences tend to exclude very low values of $\tan\beta_{1,2}$. Therefore, we
only consider
\begin{equation}
	\label{eq:3}
	\tan\beta_{1,2} > 0.3 \,,
\end{equation}
to safeguard ourselves from the constraints coming from the neutral meson mass
differences.

To deal with the constraints resulting from  $b\to s\gamma$, we
follow the procedure described in
Refs.~\cite{Florentino:2021ybj,Boto:2021qgu, Akeroyd:2020nfj}
and impose the following restriction
\begin{equation}
	\label{e:b2sg}
	2.87 \times 10^{-4} < \text{BR}(B\to X_s \gamma) < 3.77 \times 10^{-4}\,,
\end{equation}
which represents the $3\sigma$ experimental limit. As in the 2HDM, for
the case of
Type-I, this does not put strong constraints on the charged Higgs
masses.

\subsection{Perturbativity of the Yukawa couplings}

We need to ensure the perturbativity of the
Yukawa couplings. For the Type-I Yukawa structure, the top, bottom, and tau Yukawa couplings are given by
\begin{eqnarray}\label{eq:yukawa}
	y_t = \frac{\sqrt{2}\, m_t }{v \sin\beta_2}\;,
	\quad y_b = \frac{\sqrt{2}\, m_b }{v \sin\beta_2} \;,
	\quad y_\tau = \frac{\sqrt{2}\, m_\tau }{v \sin\beta_2} \;,
\end{eqnarray}
which follow from our convention that only $\phi_3$ couples to up-type
quarks, down-type quarks, and charged leptons. To maintain the
perturbativity of Yukawa couplings, we impose $\lvert y_t
	\rvert,\lvert y_b\rvert,\lvert y_\tau\rvert < \sqrt{4\pi}$.  For our
case, these constraints are all satisfied if we take into account the
lower value of $\tan\beta_2$ in~\cref{eq:3}.


\providecommand{\href}[2]{#2}\begingroup\raggedright\endgroup

\end{document}